\documentclass[twocolumn,floatfix]{revtex4}%
\usepackage[dvipdfmx]{graphicx}%
\usepackage{amsmath,amssymb}%
\usepackage{amsfonts}
\usepackage{amssymb}
\usepackage{mathrsfs}
\usepackage{color}
\usepackage{bm}
\def\d{{\partial}}
\def\s{{\sigma}}
\def\e{{\epsilon}}
\def\k{{ {\bm k} }}

\def\q{{ {\bm q} }}

\def\r{{ {\bm r} }}

\def\bmthe{{ {\bm \theta} }}
\def\bmpsi{{ {\bm \psi} }}
\def\w{{\omega}}
\def\a{{\alpha}}
\def\b{{\beta}}
\def\v{{\varphi}}
\def\g{{\gamma}}

\allowdisplaybreaks[4]

\newcommand{\maru}[1]{\raise0.2ex\hbox{\textcircled{\scriptsize{#1}}}}

\begin{document}
\title{
Quantum-metric-induced giant and reversible nonreciprocal transport phenomena in chiral loop-current phases of kagome metals
}
\author{
Rina Tazai$^{1}$, Youichi Yamakawa$^{2}$, Takahiro Morimoto$^{3}$  and Hiroshi Kontani$^2$
}
\date{\today }

\begin{abstract}
Rich spontaneous symmetry-breaking phenomena with nontrivial quantum geometric properties in metals represent central issues in condensed matter physics. In this context, the emergence of chiral loop-current order, accompanied by time-reversal symmetry-symmetry breaking in various kagome metals, has garnered significant attention. Particularly noteworthy is the giant electrical magnetochiral anisotropy (eMChA) in CsV$_3$Sb$_5$, which provides compelling evidence of time-reversal-symmetry and inversion-symmetry breakings. However, the underlying essence of this observation has remained obscured due to the lack of theoretical understanding. Here, we reveal that the loop-current order causes giant and reversible eMChA coefficient, $\gamma_{\rm eM}$, is proportional to the loop-current-induced orbital magnetization $M_{\rm orb}$ times the lifetime of conduction electrons $\tau$. In kagome metals, the derived $\gamma_{\rm eM}$ is substantial and reversible by minute magnetic fields, due to the large $\tau \ (\gg a_0/v_{\rm Fermi})$ and the field-induced reversal of $M_{\rm orb}$. By considering the experimentally observed stripe charge-density wave, the loop-current state becomes non-centrosymmetric, thereby giving rise to the eMChA. Surprisingly, the quantum-metric, which defines a fundamental geometric aspect of Bloch wavefunctions, acquires significant momentum dependence in the loop-current phase, resulting in a dramatic enhancement of eMChA by $\sim100$ times. This research not only clarifies the fundamental symmetry-breaking states in kagome metals, but also opens a new path for exploring quantum-metric-induced phenomena arising from exotic quantum phase transitions in strongly correlated metals

\end{abstract}

\address{
$^1$Yukawa Institute for Theoretical Physics, Kyoto University,
Kyoto 606-8502, Japan \\
$^2$Department of Physics, Nagoya University,
Nagoya 464-8602, Japan \\
$^3$Department of Applied Physics, The University of Tokyo, Tokyo 113-8656, Japan
}
\sloppy

\maketitle
\section{Introduction}

In the realm of condensed matter physics, the investigation of quantum phase transitions and spontaneous symmetry breaking provides fundamental insights into the essential behaviors of electrons within solids. The exploration of rotational symmetry, time-reversal symmetry (TRS), inversion symmetry (IS), and U(1) gauge symmetry-breaking quantum states has been actively pursued in various strongly correlated metals \cite{Fradkin-rev2012, Davis-rev2013, Chubukov-PRX2016, Fernandes-rev2018, Kontani-AdvPhys}.
The discovery of kagome-lattice metals $A$V$_3$Sb$_5$ ($A$=K, Rb, Cs) has paved the way for investigating novel correlation-driven quantum states and superconductivity \cite{kagome-exp1, kagome-exp2, kagome-P-Tc1, STM1, STM2, Roppongi, SC2}. Unconventional multiple-phase transitions, both with and without TRS, arise due to the suppression of conventional spin or charge density waves (DWs) by geometrical frustration. The emergence of the 2$\times$2 Star-of-David or Tri-Hexagonal DW is attributed to the triple-$\q$ bond order (BO), wherein three BOs coexist at wavevectors $\q=\q_1, \q_2, \q_3$.
This BO phase hosts intriguing chiral charge loop-current (LC) orders that violate TRS \cite{STM1, STM2}, alongside unconventional superconductivity \cite{kagome-exp1, kagome-exp2, kagome-P-Tc1, Roppongi, SC2}. Various theoretical studies have been undertaken to comprehend such TRS breaking multiple-orders \cite{Thomale2013, SMFRG, Thomale2021, Neupert2021, Balents2021, Nandkishore}. 
Theories of the BO fluctuation-mediated LC and superconductivity have been developed in Refs. \cite{Tazai-kagome,Tazai-kagome2}.

Recently, serious efforts have been devoted to the measurements of the LC order.
The staggered local magnetic field due to the LC order has been observed by several $\mu$SR studies
\cite{muSR3-Cs,muSR2-K,muSR4-Cs,muSR5-Rb}.
Importantly, the triple-$\q$ LC state has chirality, which gives rise to tiny residual uniform orbital magnetization $M_{\rm orb}$.
Therefore, the triple-$\q$ LC order is magnified under $B_z$ \cite{Tazai-Morb,Moll-hz},
and its chirality is switched by tiny $B_z$
as observed by the anomalous Hall effect (AHE) \cite{AHE1,AHE2},
nonreciprocal transport \cite{eMChA}, 
STM \cite{STM1,STM2}, 
and Kerr rotation \cite{birefringence-kagome} measurements.
Interestingly,
recent magnetic torque measurements \cite{Asaba}
reveal the emergence of the single-$\q$ LC order above $T_{\rm BO}$,
and the transition from single-$\q$ LC to triple-$\q$ LC occurs under the conical magnetic field.
Because $M_{\rm orb}=0$ in the single-$\q$ LC state due to its global TRS, 
this scenario does not conflict with previous studies
\cite{muSR3-Cs,muSR2-K,muSR4-Cs,muSR5-Rb,birefringence-kagome,eMChA,STM1}.
Notably, the LC will also have significant impact on the superconductivity, such as leading to the pair-density-wave and the $4e, 6e$ pairing states
\cite{PDW-theory,Raghu-PDW,Wu-6e,Varma-6e}.


Nonlinear effects in bulk materials provide a probe of the symmetry and geometry of Bloch electrons in solids and also useful functionalities including rectification and frequency conversion, which makes nonlinear effects important in both fundamental physics and application purposes \cite{tokura2018nonreciprocal,orenstein2021topology,ma2023photocurrent}. 
Among them, nonreciprocal transport is a current rectification effect in IS broken systems.
In particular, nonreciprocal current under the applied magnetic field is called electrical magnetochiral anisotropy (eMChA) and is intensively studied \cite{RikkenNature,RikkenBi,wakatsuki2017nonreciprocal,yokouchi2023giant}.
Recently, giant eMChA coefficient $\gamma_{\rm eM}$ in $\rho_{zz}=\rho_{zz}^0(1+\gamma_{\rm eM}B_x J_z)$ has been discovered in CsV$_3$Sb$_5$ in Ref. \cite{eMChA}.
The observation clearly indicates the breaking of IS, considering the inversion operation that transforms ${\bm j}\rightarrow -{\bm j}$ and ${\bm B}\rightarrow {\bm B}$.
The observed $\gamma_{\rm eM}$ in CsV$_3$Sb$_5$ is one or two orders of magnitude larger than
the chiral magnets, such as CrNb$_3$S$_6$ \cite{CrNb3S6} and MnSi \cite{MnSi}, 
in which the chiral spin texture gives the eMChA.
Unfortunately, the fundamental electronic states of the LC phase indicated by this important experiment are still unclear because of the lack of the underlying microscopic mechanism for the giant and swithcable eMChA in CsV$_3$Sb$_5$.


The observed giant eMChA and AHE manifest that kagome metal is an ideal platform for developing novel quantum geometric effects
\cite{resta94,marzari,resta2011insulating,nagaosa2010anomalous,chiu16,ma23}.
Nontrivial geometry of the Bloch electrons in solids is captured by the quantum geometric tensor ${\cal T}=G- i\Omega/2$ that measures the distance of the Bloch wave functions for different momenta in the Hilbert space 
\cite{marzari,resta2011insulating}.
The imaginary part $\Omega$ is the Berry curvature that gives rise to the anomalous (or spin) Hall effects \cite{Kontani-ROP,Kontani-SHE-PRL,Kontani-Yamada,nagaosa2010anomalous}
and nonlinear Hall effect \cite{sodemann2015quantum}.
The real part $G$ is the quantum metric (QM) and characterizes, for example, wave functions in flat bands, superfluid weight \cite{torma2022superconductivity}, and nonlinear transport \cite{gao2023quantum}.
In kagome metals, $\Omega$ causes the large anomalous Hall effect
\cite{Tazai-kagome2,AHE-kagome-theory} and orbital magnetization \cite{Tazai-Morb}, while the impact of the QM in kagome metals is still uncovered.
Importantly, in kagome metals, nonvanishing ${\cal T}$ originates from the BO and LC: Their microscopic mechanisms due to strong electron correlations and geometrical frustrations have been developed in Refs.  
\cite{Thomale2013,SMFRG,Thomale2021,Neupert2021,Balents2021,Tazai-kagome,Tazai-kagome2,Zhou-cLC-RPA,Nat-g-ology}.
Beyond-mean-field electron correlations give rise to various BO and LC orders in cuprates and Fe-based superconductors
\cite{Fradkin-rev2012,Davis-rev2013,Chubukov-PRX2016,Fernandes-rev2018,Kontani-AdvPhys,Varma,Nersesyan,Onari-SCVC,Tsuchiizu1,Tsuchiizu4,Yamakawa-Cu,Yamakawa-FeSe,Onari-FeSe,Tazai-Matsubara,Tazai-cLC,Tazai-rev2021}
and twisted bilayer graphene
\cite{Chubukov-gr,Onari-TBG}.

In this study, we investigate the mechanism of the field-tunable giant eMChA arising from chiral LC in kagome metals, by focusing on the $2a_0$ or $4a_0$ ($a_0$: the unit cell length) stripe charge DW (CDW) that is universally observed in Cs-based kagome metals by STM and NMR measurements \cite{stripe1, stripe2}. 
The realized electronic states, characterized by TRS and IS breaking, lead to a giant eMChA coefficient $\gamma_{\rm eM} \propto M_{\rm orb} \tau I_{\rm st}$, where $I_{\rm st}$ is the stripe CDW potential and $\tau$ is the lifetime of conduction electrons. 
In kagome metals, the derived eMChA becomes substantial due to large $\tau \ (\gg a_0/v_{\rm Fermi})$, and exhibits a sharp change in sign upon variation of the small magnetic field $B_z$, consistent with observations in Cs-based kagome metals. 
Unexpectedly, the QM acquires significant momentum dependence in the loop-current phase, resulting in a dramatic enhancement of the eMChA by $\sim100$ times.
This mechanism dominates the previously studied Berry-curvature dipole mechanism ($\gamma_{\rm eM}\propto\tau^0$) and the QM dipole mechanism ($\gamma_{\rm eM}\propto\tau^{-1}$) in kagome metals with large $\tau$.
The present theory not only explains the switchable giant eMChA in the LC phase, which is a key experiment in kagome metals, but also reveals the emergence of resonantly-enlarged QM that will cause various exotic phenomena.

\section{Introduction of kagome lattice models}
\label{sec:Kagome-lattice-model}

The unit cell of the original kagome lattice model is
composed of the three sublattices A, B, C shown in Fig. \ref{fig:fig1} (a).
Under the triple-$\q$ density-waves, the unit cell is enlarged to 
twelve sublattices ($l=1-12$), as denoted in Figs. \ref{fig:fig1} (a) and (b).
The Fermi surfaces are mainly composed of $d_{xz}$ and $d_{yz}$ orbitals denoted in Figs. \ref{fig:fig1} (b).
The $d_{xz}$ orbital mainly composes the ``pure-type'' Fermi surface depicted in Fig. \ref{fig:fig1} (c),
where each van-Hove singular (vHS) point $\k\approx \k_{X}$ ($X=$A, B, C) is composed of single sublattice $X$, called as ``sublattice interference'' \cite{Thomale2013}.
Both BO and LC orders emerge on $d_{xz}$-orbitals due to strong electron correlations \cite{Tazai-kagome,Tazai-kagome2}.


\subsection{2D kagome lattice model: LC+BO order parameters}

\begin{figure}[htb]
\includegraphics[width=.99\linewidth]{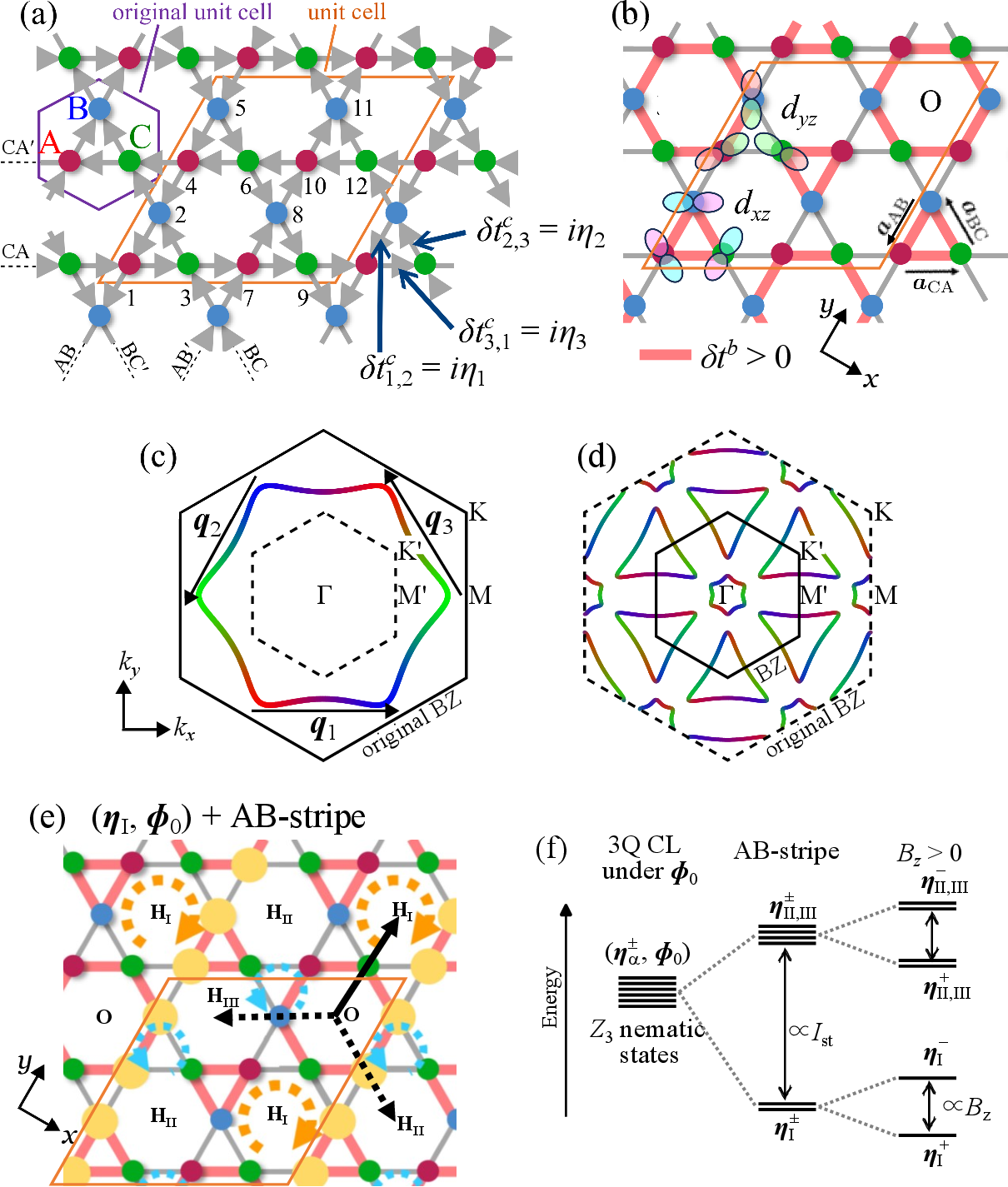}
\caption{
{Lattice structure, Fermi surface, and BO form factor in kagome metal}: \
(a) 2D kagome lattice model with $2\times2$ triple-$\q$ LC 
${\bm \eta}=(\eta_1,\eta_2,\eta_3)$ on $d_{xz}$ orbital.
Note that $(\eta_1,\eta_2,\eta_3)$ and $(\eta_1,-\eta_2,-\eta_3)$
are the same bulk states after shifting ${\bm R}={\bm a}_{\rm BC}+{\bm a}_{\rm CA}$.
The extended unit cell contains twelve sublattices $l=1\sim12$.
(b) Triple-$\q$ Tri-Hexagonal BO ${\bm \phi}=\phi{\bm e}_0$ $(\phi>0)$
on $d_{xz}$ orbital. 
Two orbitals ($d_{xz}$, $d_{yz}$) on the $V$-ion are depicted.
(c) Original Fermi surfaces without density waves. 
(d) Folded Fermi surfaces under the $2\times2$ density waves with the folded BZ. 
(e) Nematic LC+BO coexisting state $({\bm\eta}_{\rm I},{\bm\phi}_0)$.
The $C_6$ symmetry centers of the LC and BO are expressed as ${\rm O}$ and ${\rm H}_{\rm I}$, respectively.
The director is parallel to ${\rm O}$-${\rm H}_{\rm I}$ line.
(${\rm H}_{\rm II(III)}$ is the center of ${\bm\eta}_{\rm II(III)}$.)
The $2a_0$ stripe CDW along AB direction is shown by the yellow circles.
(f) Lift of the six-fold degenerated nematic states $({\bm\eta}_\a^\pm,{\bm\phi}_0)$ ($\a=$I, \ II, \ III) due to the stripe potential and the magnetic field.
}
\label{fig:fig1}
\end{figure}

To begin with, we introduce the $2\times2$ density-waves in the 2D kagome lattice model.
The folded pure-type Fermi surfaces are shown in Fig. \ref{fig:fig1} (d) together with the folded Brillouin zone (BZ).
The three wavevectors are given as
$\q_1= {\bm a}_{\rm AB}\times{\hat z}/S$,
$\q_2= {\bm a}_{\rm BC}\times{\hat z}/S$, and 
$\q_3= {\bm a}_{\rm CA}\times{\hat z}/S$, 
where $S=\sqrt{3}a_0^2/2\pi$ is the area of the original unit cell divided by $\pi$.
(Here, $a_0=2|{\bm a}_{\rm AB}|$.) 
The form factor (=normalized $\delta t_{ij}^{\rm c}$)
with ${\bm q}={\bm q}_1$, $f_{ij}^{(1)}$, is 
$+i$ when the sites $(i,j)$ belongs to the sublattices $(l,m)=(1,2),(2,4),(4,5),(5,1)$, 
and $-i$ for $(7,8),(8,10),(10,11),(11,7)$.
Importantly, the odd parity relation $f_{ij}^{(1)}=-f_{ji}^{(1)}$ holds for the LC.
Other form factors with ${\bm q}_2$ and ${\bm q}_3$, 
$f_{ij}^{(2)}$ and $f_{ij}^{(3)}$, are also derived from Fig. \ref{fig:fig1} (a).
Using the notation ${\bm f}_{ij}=(f_{ij}^{(1)},f_{ij}^{(2)},f_{ij}^{(3)})$, 
the LC form factor is given as
\begin{eqnarray}
\delta t_{ij}^{\rm c}&=& {\bm \eta}\cdot{\bm f}_{ij} ,
\label{eqn:tc}
\end{eqnarray}
%
where
${\bm \eta}\equiv(\eta_1,\eta_2,\eta_3)$ is the set of 
the current order parameters with the wavevector $\q_n$ ($n=1,2,3$).
Note that the LC form factors are Hermitian $\delta t_{ij}=(\delta t_{ji})^*$,
which leads to the relation $f_\q^{lm}(\k)=(f_{-\q}^{ml}(\k+\q))^*$
\cite{Kontani-AdvPhys}.

Also, the BO form factor shown in Fig. \ref{fig:fig1} (b) is given as
\begin{eqnarray}
\delta t_{ij}^{\rm b}&=& {\bm \phi}\cdot{\bm g}_{ij} ,
\label{eqn:tb}
\end{eqnarray}
where ${\bm \phi}\equiv(\phi_1,\phi_2,\phi_3)$ is the set of the BO parameters
with the wavevector $\q_n$ ($n=1,2,3$),
and $g_{ij}^{(n)}=g_{ji}^{(n)}=+1$ or $-1$ as shown in Fig. \ref{fig:fig1} (b).
For example, $g_{ij}^{(1)}=+1 \ [-1]$ for the sites $(i,j)$ that belong to the sublattices $(1,2),(4,5),(8,10),(11,7)$ [$(2,4),(5,1),(7,8),(10,11)$].

Here, we introduce the notation ${\bm e}_0=(1,1,1)/\sqrt{3}$, ${\bm e}_{\rm I}=(1,-1,-1)/\sqrt{3}$, ${\bm e}_{\rm II}=(-1,1,-1)/\sqrt{3}$, ${\bm e}_{\rm III}=(-1,-1,1)/\sqrt{3}$.
In the following, we fix the triple-$\q$ BO as ${\bm\phi}_0\equiv \phi{\bm e}_0$ without loss of generality.
The triple-$\q$ LC is expressed as ${\bm\eta}_\a\equiv \eta{\bm e}_\a$, where $\a=0$ or $\a={\rm I, \ II, \ III}$.
In the case of $\a={\rm I-III}$, the $C_6$ symmetry center of the LC, denoted as ${\rm H}_\a$, differs from that of the BO, denoted as ${\rm O}$.
This out-of-phase LC+BO coexisting state $({\bm\eta}_\a,{\bm\phi}_0)$ exhibits $Z_3$ nematicity, with its director aligned along the ${\rm O}$-${\rm H}_\a$ line \cite{Tazai-kagome2}.
The LC+BO state for $\a={\rm I}$ is shown in Fig. \ref{fig:fig1} (e).

In this paper, we study the eMChA in the $Z_3$ nematic state, which is realized when $T_{\rm LC} < T_{\rm BO}$, according to the Ginzburg-Landau (GL) free-energy theory \cite{Tazai-kagome2,Tazai-Morb}.
 A brief explanation is also provided in Appendix A.
It is important to emphasize that the eMChA occurs as long as the LC and stripe CDW are present, even in the absence of the BO. 
However, we discuss the LC+BO coexistence in Sect. \ref{sec:stripe-zeeman} to understand the field dependence of the LC order parameter.

\subsection{Stripe CDW, Zeeman splitting}
\label{sec:stripe-zeeman}


The $Z_3$ degeneracy of the LC+OB states is easily lifted by additional potentials and fields.
By symmetry, it is lifted as $E_{\rm I}\ne E_{\rm II(III)}$ by introducing the $2a_0$ stripe CDW,
depending on the angle between the director of $Z_3$ nematicity and the stripe CDW.
(The sign of $E_{\rm I}-E_{\rm II(III)}$ would depend on the model parameters.)
Figure \ref{fig:fig1} (e) exhibits the state $({\bm\eta}_{\rm I},{\bm\phi}_0)$
under the AB-direction stripe.

The degeneracy is further lifted by introducing the magnetic field due to orbital-Zeeman energy $-M_{\rm orb}B_z$,
where the orbital magnetization is expressed as
$M_{\rm orb}[{\bm \eta}_\a^\pm,{\bm\phi}_0] = \pm \eta[-m_1\phi/3+m_2\eta^2/(3\sqrt{3})-m_3\phi^2/(3\sqrt{3})]$ for any $\a={\rm I, \ II, \ III}$,
as we discuss in Ref. \cite{Tazai-Morb} and Appendix A.
The coefficients $m_1,m_2,m_3$ are derived in Ref. \cite{Tazai-Morb}.
Here, ${\bm\eta}_\a^\pm\equiv \pm\eta{\bm e}_\a$ with $\eta>0$.
Therefore, ${\bm\eta}_{\a}^+$ is switched to ${\bm\eta}_{\a'}^-$ under the magnetic field $B_z$, while $|\eta|$ is unchanged.
The result is summarized in Fig. \ref{fig:fig1} (f).
(The magnitudes of $\phi$ and $\eta$ would be unchanged by small ${\bm B}$.)


As we will explain later,
the eMChA conductivity under the AB direction stripe CDW 
is proportional to the first component of the LC order;
$[{\bm\eta}_{\rm I}^\pm]_1\propto \pm\eta$ for $({\bm\eta}_{\rm I},{\bm\phi}_0)$ and
$[{\bm\eta}_{\rm II(III)}^\pm]_1\propto\mp\eta$ for $({\bm\eta}_{\rm II(III)},{\bm\phi}_0)$.
In both cases, the eMChA is switched by the direction of $B_z$,
which is consistent with experimental observations.

\subsection{Three-dimensional LC stacking states}

Here, we consider the three-dimensional (3D) LC state.
The $2\times2\times2$ triple-$\q$ LC is expressed by
${\bm\eta}={\bm\eta}_\a$ for the first layer
and the set of the inter-layer phase shifts 
${\bmthe}=(\theta_1,\theta_2,\theta_3)$ with $\theta_i=0$ or $\pi$.
Then, the second layer order is ${\bm\eta}'=(e^{i\theta_1}\eta_1,e^{i\theta_2}\eta_2,e^{i\theta_3}\eta_3)$.
In this paper, we consider the following two cases
because they are stabilized in the LC+BO coexisting state
\cite{Balents2021,Tazai-kagome2,Tazai-Morb}.
(1) ${\bmthe}=(0,0,0)$, where the same 2D orders are stacked vertically,
which we call the ``vertical LC (v-LC)'' state. (See Fig. \ref{fig:fig2} (e).) 
(2) ${\bmthe}=(0,\pi,\pi)$, where the 2D order on even-number layers and that on the odd-number layers are shifted by the vector ${\bm a}_{\rm AB}$, which we call the ``shift-stack LC (s-LC)'' state. (See Fig. \ref{fig:fig2} (g).)
The director of the nematicity of ${\bmthe}=(0,\pi,\pi)$ s-LC state is along ${\bm a}_{\rm AB}$.
In Appendix A, we discuss that the v-LC (s-LC) will appear inside the v-BO (s-BO) phase by focusing on the $\eta^2\phi$-order term of the GL free energy.
Once IS is broken, the resulting eMChA conductivities for the v-LC and s-LC become nearly identical.

\section{Inversion symmetry breaking in LC+stripe state}
\label{sec:ISB}

In this section, we explain that IS is violated by introducing the $2a_0$ stripe CDW into both v-LC and s-LC states.
The stripe CDW is generally observed in Cs-based kagome metals
\cite{stripe1,stripe2}.
The absence of IS is a necessary condition for nonlinear conductivities.

\subsection{Inversion symmetry breaking in 2D kagome lattice model}

\begin{figure*}[htb]
\includegraphics[width=.75\linewidth]{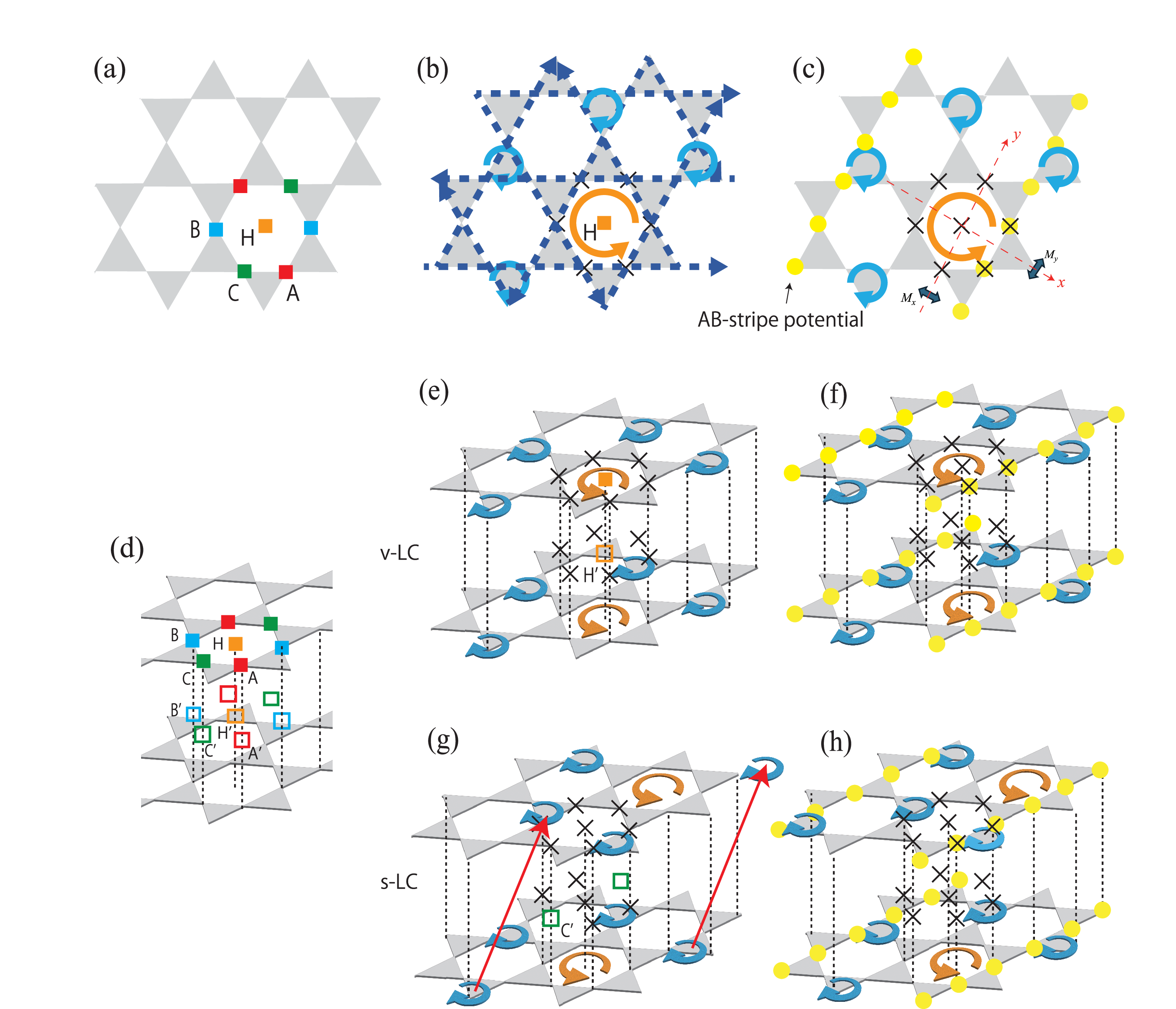}
\caption{
{Inversion symmetry breaking in LC+stripe state}: \
(a) Original 2D crystal structure. The IS centers are described by A, B, C, H. 
(b) $2\times2$ LC order with remaining IS center H. 
(c)  Coexisting phase of $2\times2$ LC and $2a_0$ stripe CDW. 
Thus, IS is broken when the LC and stripe CDW coexist. 
(d) Original 3D crystal structure. The IS centers are described by A, B, C, H, A', B', C', H'. 
(e) v-LC with remaining IS centers H and H'. 
(f) Coexisting phase of v-LC and $2a_0$ stripe CDW. 
In this case, IS is broken. 
(g) s-LC with remaining IS center C'. 
(h)  Coexisting phase of s-LC and $2a_0$ stripe CDW,
where the shift-stacking direction and the stripe direction are parallel. 
In this case, IS is broken. 
}
\label{fig:fig2}
\end{figure*}

In the original 2D kagome lattice shown in Fig. \ref{fig:fig2} (a),
the IS centers are located at A, B, C, and H,
represented by filled square symbols.
Note that H is the center of the $C_6$ rotational symmetry.
When the triple-$\q$ LC order is introduced, only the inversion center H survives as shown in Fig. \ref{fig:fig2} (b).
Note that the inversion center H survives even if the triple-$\q$ LC and 
triple-$\q$ BO coexist, in both in-phase ($C_6$ symmetry) and out-of-phase ($C_2$ symmetry) states
\cite{Tazai-kagome2}.
Thus, IS is not violated by triple-$\q$ LC and triple-$\q$ BO.
We find that IS is broken by introducing the stripe CDW,
as in Fig \ref{fig:fig2} (c).
Here, the yellow circles represent the stripe CDW potentials $I_{\rm st}$,
which is consistent with the energy-independent (vertical) 
QPI signal with period $2a_0$ observed in Ref. \cite{stripe1,stripe2}.
Hereafter, we analyze only the $2a_0$ stripe CDW in the numerical study.
(IS is also violated by introducing the period $4a_0$ CDW,
while stripe CDW alone does not break IS.)

\subsection{Inversion symmetry breaking in 3D kagome lattice model}

In the original 3D crystal structure shown in Fig. \ref{fig:fig2} (d),
the IS centers are located at A, B, C, H
(filled square symbols), and A', B', C', H' (open square symbols).
Here, X' are the middle points between two layers.
Figure \ref{fig:fig2} (e) shows a v-LC state with ${\bmthe}=(0,0,0)$.
Here, only the IS centers H and H' survive as shown in Fig. \ref{fig:fig2} (e).
In Fig. \ref{fig:fig2} (f), we introduce the $2a_0$ stripe CDW.
Then, none of the IS centers remains as in Fig. \ref{fig:fig2} (f).
Then, we find that the eMChA becomes finite.
Figure \ref{fig:fig2} (g) shows the s-LC state with ${\bmthe}=(0,\pi,\pi)$.
Here, the triple-$\q$ LC on the even-number layers and that on the odd-number layers are shifted by the vector ${\bm a}_{\rm AB}$.
In this $2\times2\times2$ s-LC state,
only the IS center C' survives
as shown in Fig. \ref{fig:fig2} (g).
By introducing the $2a_0$ stripe CDW along AB direction, 
none of the IS centers remain as in Fig. \ref{fig:fig2} (h).
Then, we find that the eMChA becomes finite.
Note that the IS center C' remains intact if the $2a_0$ stripe CDW is along BC or CA directions, and therefore the eMChA disappears.

To summarize, 
by introducing the $2a_0$ stripe CDW that is generally observed in Cs-based compounds,
IS is broken in both 
the v-LC state in Fig. \ref{fig:fig2} (f) ($\bmthe=(0,0,0)$)
and the s-LC state in Fig. \ref{fig:fig2} (h) ($\bmthe=(0,\pi,\pi)$).
In both cases, finite eMChA is expected to be realized.
(In contrast, the alternative v-LC state with $\bmthe=(\pi,\pi,\pi)$ preserves IS even after introducing the stripe, so the eMChA vanishes.)
It is noteworthy that IS in the single-$\q$ LC state is broken 
by introducing the stripe CDW that is parallel to the single-$\q$ LC order.
Therefore, the eMChA can appear even in the single-$\q$ LC state.

\section{NLH and eMChA conductivities} 

In this section, we study the second-order response to the electric field ${\bm E}$ in the LC state in kageme metals.
The nonlinear Hall conductivity $\s_{\a\b\b'}$ is defined as
\begin{eqnarray}
\Delta j_\a=\s_{\a\b\b'}E_\b E_{\b'} ,
\end{eqnarray}
where $j_\a$ is the charge current along the $\a$-axis of order $O(E^2)$.
Here, $\a,\b,\b'=x$, $y$, or $z$.
The eMChA conductivity $\s_{\a\b\b',\g}$ is defined as
\begin{eqnarray}
\Delta j_\a=\s_{\a\b\b',\g}E_\b E_{\b'}B_\g ,
\label{eqn:s_abcd}
\end{eqnarray}
where $B_\gamma$ is the magnetic field along the $\gamma$ axis.
Under the inversion operation, $\Delta{\bm j}$ and ${\bm E}$ are reversed 
while ${\bm B}$ is unchanged.
In addition, $\s_{\a\b\b'}$ and $\s_{\a\b\b',\g}$ 
are unchanged if the system is centrosymmetric.
Therefore, $\s_{\a\b\b'}=\s_{\a\b\b',\g}=0$ in any centrosymmetric system.
Thus, the absence of IS is a necessary condition for the finite 
$\s_{\a\b\b'}$ and $\s_{\a\b\b',\g}$.
Note that the eMChA coefficient is given as $\gamma_{\rm eM}\equiv-\s_{zzz,x}/(\s_{zz})^2$.
Under the time-reversal operation,
the signs of $\Delta{\bm j}$ and ${\bm B}$ are reversed.
For this reason, the eMChA (NLH) conductivity is $T$-even ($T$-odd).

\subsection{Symmetry analysis of the realization conditions for eMChA}
\label{sec:Symmetry}

To begin, we discuss the realization conditions for eMChA in kagome metals from the perspective of symmetry.
In the LC+stripe state shown in Fig. \ref{fig:fig2} (c),
$M_{x}$ ($M_y$) represents the mirror operation $x\rightarrow-x$ ($y\rightarrow-y$)
and the inversion operation is $I=M_x M_y M_z$.
Below, we assume $2\times2\times1$ LC to simplify the discussion. 
We denote the transformation of a physical quantity $A$ under the symmetry operation $X$ as $A^X$.

For $X=I$, real coordinate $\r$, charge current ${\bm j}$, and magnetic field ${\bm B}$ are respectively transformed as $\r^I=-\r$, ${\bm j}^I=-{\bm j}$, and ${\bm B}^I={\bm B}$.
Note that $\r$ and ${\bm j}$ are polar vectors, while ${\bm B}$ is an axial vector.
For $X=M_x$, we obtain $\r^{M_x}=(-x,y,z)$, ${\bm j}^{M_x}=(-j_x,j_y,j_z)$, and ${\bm B}^{M_x}=(B_x,-B_y,-B_z)$.
Next, we consider the transformation of the model order parameters shown in Fig. \ref{fig:fig2} (c), where the LC order is ${\bm\eta}=\eta{\bm e}_{\rm I}$ and the stripe CDW potential denoted by the yellow circles is $I_{\rm st}$. ($-I_{\rm st}$ on the other A, B sites.)
Then, the following relations are derived from Fig. \ref{fig:fig2} (c):
$(\eta, I_{\rm st})^{M_x}=(-\eta, -I_{\rm st})$,
$(\eta, I_{\rm st})^{M_y}=(-\eta, I_{\rm st})$, and 
$(\eta, I_{\rm st})^{T}=(-\eta, I_{\rm st})$, 
where $T$ is the time-reversal operation.
The symmetries of model order parameters are summarized in Table \ref{tab:tab2}.

\begin{table} 
\caption{\label{tab:tab2}
Parity of each model parameter ($\eta, \phi, I_{\rm st}$) under the operation $X=M_x$, $M_y$, $I$, and $T$.
Here, the BO parameter is $\phi{\bm e}_0$.
} 
\begin{ruledtabular} 
\begin{tabular}{|c||c|c|c|}
\hline
 & $\eta$ & $\phi$ & $I_{\rm st}$ \\ 
\hline
$M_x$ & $-1$ & $1$ & $-1$  \\
\hline
$M_y$ & $-1$ & $1$ & $1$  \\
\hline
$I$ & $1$ & $1$ & $-1$  \\
\hline
$T$ & $-1$ & $1$ & $1$  \\
\hline
\end{tabular} 
\end{ruledtabular} 
\end{table}

When $\eta I_{\rm st}\ne0$, the spin-independent band dispersion of the present model loses its inversion symmetry, resulting in $\e_{b,\k}\ne\e_{b,-\k}$.
In more detail, the relations $\e_{b,k_x,k_y}=\e_{b,-k_x,k_y}\ne \e_{b,k_x,-k_y}$ is satisfied because $[\eta I_{\rm st}]^{M_{x}}=\eta I_{\rm st}$ and $[\eta I_{\rm st}]^{M_{y}}=-\eta I_{\rm st}$.
Therefore, the Fermi surfaces shift along the $k_y$-axis as indicated by Fig. \ref{fig:fig3} (b), due to the dipole potential
$\e_{b,\k}^-\equiv \frac12 (\e_{b,\k}-\e_{b,-\k})$; see Sect. \ref{sec:Dipole} for more detail.

Here, we discuss the symmetry of the eMChA conductivity given by Eq. (\ref{eqn:s_abcd}).
Under the symmetry operation $X$, it is transformed as 
\begin{eqnarray}
\Delta j_\a^X=[\s_{\a\b\b',\g}]^X E_\b^X E_{\b'}^X B_\g^X ,
\label{eqn:s_abcd2}
\end{eqnarray}
where $[\s_{\a\b\b,\g}]^X$ obeys the same symmetry transformation as that for the Hamiltonian.
When $X=I$, Eq. (\ref{eqn:s_abcd2}) gives $-\Delta j_\a=[\s_{\a\b\b',\g}]^I E_\b E_{\b'} B_\g$.
Therefore, IS breaking ({\it i.e.}, $[\s_{\a\b\b',\g}]^I \ne\s_{\a\b\b',\g}$) is a necessary condition for the emergence of the eMChA.
Now, we set $\a=\b=\b'=z$ and $\gamma=x$ to understand the experimentally observed eMChA in kagome metals.
When $X=M_y$, Eq. (\ref{eqn:s_abcd2}) gives $\Delta j_z=[\s_{zzz,x}]^{M_y} E_z^2 (-B_x)$.
Therefore, $\s_{zzz,x}$ can appear when $M_y$ symmetry is broken, as observed in the present model from Table \ref{tab:tab2}; $[\eta I_{\rm st}]^{M_y}=-\eta I_{\rm st}$.

When $X=T$, Eq. (\ref{eqn:s_abcd2}) gives $-\Delta j_z=[\s_{zzz,x}]^{T} E_z^2 (-B_x)$, where
$[\s_{zzz,x}]^{T}=(-1)^{m+m'}\s_{zzz,x}$ when $\s_{zzz,x}\propto \tau^m \eta^{m'}$ with the relaxation time $\tau$.
Therefore, $\s_{zzz,x}$ can appear when $m+m'$ is even.
In Boltzmann theory, the most divergent eMChA conductivity with respect to $\tau \ (\gg a_0/v_{\rm Fermi})$ is $O(\tau^3)$, as we explain in Sect. \ref{sec:Boltzmann} and Appendix C.
Thus, giant $\tau^3$-eMChA conductivity can appear for odd $m'$, especially for $m'=1$ in the least order.

To summarize, the giant and reversible eMChA conductivity $\s_{zzz,x}\propto \tau^3 \eta$ in the LC phase of kagome metals is understandable from the perspective of symmetry.
Below, we conduct a numerical study of eMChA conductivity to gain a quantitative understanding.

\subsection{$d_{xz}$+$d_{yz}$ orbital 3D kagome metal model}
\label{sec:dxz-dyz}

\begin{figure*}[htb]
\includegraphics[width=.7\linewidth]{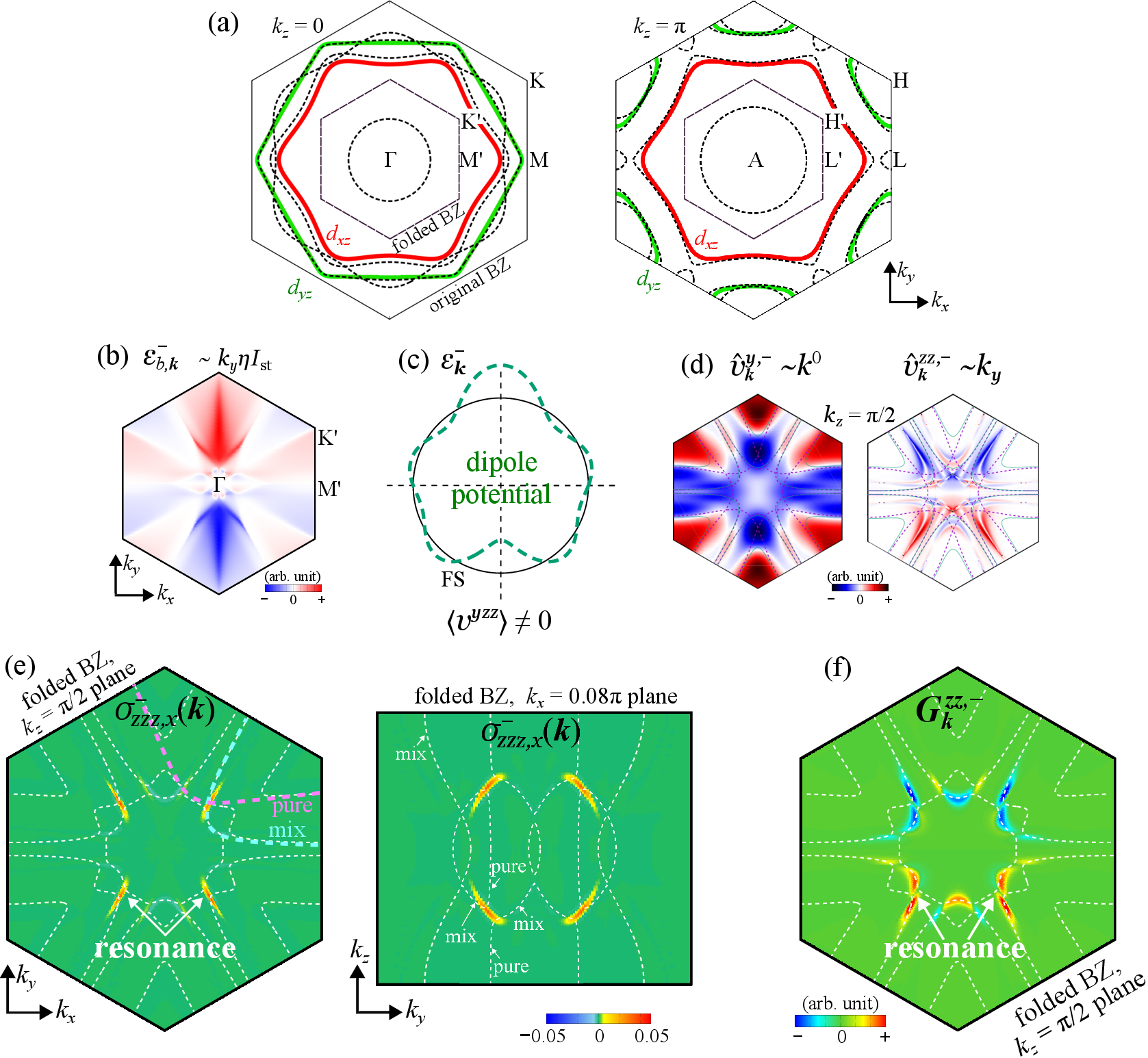}
\caption{
{eMChA due to dipole potential and qQM-induced enhancement}: \
(a) Fermi surfaces of the $d_{xz}+d_{yz}$ orbital 3D kagome lattice model in the original BZ.
The $d_{xz}$ orbital weight is expressed as red color. 
These Fermi surfaces qualitatively fit the main Fermi surfaces of the first-principles DFT calculation (brown dotted lines). 
(b) Dipole potential $\e_{b,\k}^{-}$ ($b=9$) in the folded BZ for $\eta=0.05$ and $I_{\rm st}=0.01$. 
(c) Schematic picture of the dipole potential.
It gives no macroscopic current ($\langle v \rangle={\bm0}$) because of the Bloch's theorem.
However, it gives finite $\langle v^{yzz} \rangle$.
Then, the NLH $\s_{yzz}$ and the eMChA $\s_{zzz,x}$ becomes finite. 
(d) ${\hat v}_\k^{y,-}$ and ${\hat v}_\k^{zz,-}$ on the Fermi surface in the folded BZ.
They originate from $\e_{b,\k}^{-}$ in (b). 
(e) $\s_{zzz,x}^{-}(\k)$ in $k_z=\pi/2$ plane and $k_x=0.08\pi$ plane
for $\eta=0.024$ and $I_{\rm st}=0.01$ at $T=0.01$.
It is an even function of $\k$, so its $\k$ summation remains finite.
(f) $\eta$-odd part of the QM at the Fermi level, $G^{zz,-}_\k$, in the $\k_z=\pi/2$ plane. 
Its resonant positions are the same as those of $\s_{zzz,x}^{-}(\k)$ in (e). 
Thus, giant eMChA originates from the resonantly enhanced $\eta$-odd QM.
}
\label{fig:fig3}
\end{figure*}

To perform a quantitative numerical analysis of the nonlinear conductivities in kagome metals,
we introduce a 3D kagome lattice model with $d_{xz}$ and $d_{yz}$ orbitals.
As we explained in Sect. \ref{sec:Kagome-lattice-model},
the pure-type Fermi surface is mainly composed of the $d_{xz}$ orbital, on which the LC order emerges.
In addition, the $d_{yz}$ orbital mainly composes the ``mix-type'' Fermi surface, where each vHS point is made of two sublattices.
These Fermi surfaces of the present 3D kagome metal are shown in Fig. \ref{fig:fig3} (a) for the electron filling at $n_{\rm exp}=4.6$.
The $d_{xz}$ orbital weight is expressed as red color. 
The inter-layer hopping integrals for $d_{yz}$ orbital is introduced
to reproduce the $k_z$-dependence of the vHS energy of the mix-type band 
\cite{ARPES-VHS,ARPES-band}.
These Fermi surfaces are qualitatively consistent with
the main Fermi surfaces of the first-principles DFT calculation for CsV$_3$Sb$_5$, which are shown by brown dotted lines.
The model parameters and the bandstructure are explained in Appendix B.

In kagome metals, both pure-type and mix-type Fermi surfaces are very close in energy on the Fermi surfaces.
In addition, the DFT calculation shows that
the pure/mix-type Fermi surface contains $10\sim30$\% $d_{yz}$/$d_{xz}$-orbital component \cite{Tazai-Morb}.
Therefore, the $2\times2$ LC order on the $d_{xz}$ orbitals, ${\bm\eta}_\a$, causes prominent pure-mix band hybridization and the Fermi surface reconstruction.
This LC-induced band reconstruction causes not only large $M_{\rm orb}$
\cite{Tazai-Morb} but also giant eMChA as we will explain below.


\subsection{Dipole potential without TRS due to finite $\eta I_{\rm st}$}
\label{sec:Dipole}

In the triple-$\q$ LC state, IS is broken by introducing a $2a_0$ or $4a_0$ CDW potential, as explained in Sect. \ref{sec:ISB}.
Here, we specifically introduce the $2a_0$ stripe CDW potential $I_{\rm st}$, as schematically illustrated in Fig. \ref{fig:fig2} (c).
In this paper, we denote the $\eta$-odd and $\eta$-even parts of the function $A$ as $A^{-}$ and $A^{+}$, respectively.
They are defined as 
\begin{eqnarray}
A_{\k,\eta}^\pm\equiv \frac12 ([A]_{\k,\eta} \pm [A]_{\k,-\eta}).
\end{eqnarray}
%
In the following, we frequently refer to $\e_{b,\k,\eta}^\pm\equiv \frac12 (\e_{b,\k,\eta} \pm \e_{b,\k,-\eta})$.
In the present model, the relation $\e_{b,\k}^\pm = \frac12 (\e_{b,\k} \pm \e_{b,-\k})$ is satisfied for a fixed $\eta$,
because time-reversal operation gives $\e_{b,\k,\eta}=\e_{b,-\k,-\eta}$ for a fixed $I_{\rm st}$.
Therefore, $\e_{b,\k}^-$ is both the ``$\eta$-odd part'' and the ``$\k$-odd part'' of $\e_{b,\k}$.

Figure \ref{fig:fig3} (b) shows the numerical result of a conduction band $\e_{b,\k}^-$ ($b=9$) under the AB stripe CDW for $\eta=0.05$.
$\e_{b,\k}^-$ give a ``dipole potential in $\k$-space'' because it is proportional to $k_y$.
It has the symmetry-protected nodes on the $k_x$-axis line and other two nodes on $k_y\approx\pm \sqrt{3} k_x$ line.
Its schematic picture is given in Fig. \ref{fig:fig3} (c).
To understand the origin of the dipole potential,
we consider the symmetry of the triple-$\q$ LC order depicted in Fig. \ref{fig:fig1} (a) and Fig. \ref{fig:fig2} (b).
The kagome lattice belongs to the $D_{6h}$ point group.
Around point $H$, the LC pattern preserve IS,
while all the parities of the six mirror operations along the $\theta=(\pi/6)m$ axis ($m=0,1,\cdots5$ for $3\s_h$ and $3\s_v$) are $-1$.
Thus, the LC order belongs to the $A_{2g}$ irreducible representation,
whose basis function is $b_{A_{2g}}=x(x^2-3y^2)y(3x^2-y^2)$.
Also, the AB direction stripe CDW changes its sign under $M_x$ mirror operation, 
while it is unchanged under $M_y$ mirror operation.
Based on the second-order perturbation theory, the derived $\e_\k^-$ of order $O(\eta I_{\rm st})$ is proportional to $x \times b_{A_{2g}}\sim k_y(k_x^2-3k_y^2)(3k_x^2-k_y^2)$, which reproduces the $k_y$-dipole with complex sign reversals shown in Fig. \ref{fig:fig3} (b).
Thus, the dipole potential is simply expressed as $\e_\k^-\propto k_y \eta I_{\rm st}$ with accidental sign reversals as shown in Figs. \ref{fig:fig3} (b) and (c).

Due to the dipole potential $\e_\k^-$,
one may consider that the bulk current $\langle v^y \rangle\propto \sum_\k v^y f$ is finite, 
where $f$ is the Fermi distribution function, which is expanded as
$f(\e_\k)\approx f(\e_\k^+)+f'(\e_\k^+)\e_\k^-$.
However, $\langle v^y \rangle$ is exactly zero because the LC order is periodic in the $2\times2$ unit cell, known as the Bloch's theory based on the Galilean transformation argument \cite{Bloch}.
This is also simply proved as $\langle {\bm v} \rangle\propto \sum_\k {\bm v} f(\e_\k)=\sum_\k {\bm\nabla}_\mu(\e_\k-\mu) f(\e_\k)=-\sum_\k (\e_\k-\mu)f'(\e_\k){\bm v}={\bm0}$ at $T=0$.

\subsection{Boltzmann transport theory and quantum metric}
\label{sec:Boltzmann}


Here, we focus on the most divergent linear and nonlinear conductivities with respect to the $\tau$ based on the relaxation time approximation.
Then, $\s_{\a\a}$, $\s_{\a\b\b}$, and $\s_{\a\b\b,\g}$ are respectively given by $f_1$, $f_2$, and $f_3$,
where $f_n$ ($n\ge1$) represents the nonequilibrium part of the distribution function in proportion to $\tau^n$.
The detailed analysis is performed in Appendix C.
The derived formulae are given as
\begin{eqnarray}
\s_{\a\a}&=& -e^2 \frac{g_s}{N}\sum_{a,\k} (f' v^\a v^\a)_{a,\k}\frac1{2\gamma}
\label{eqn:sxx} \\
\s_{\a\b\b}&=& e^3 \frac{g_s}{N}\sum_{a,\k} (f' v^\b v^{\a\b})_{a,\k}\frac1{4\gamma^2},
\label{eqn:NLH} \\
\s_{\a\b\b,\gamma}
&=& e^4 \frac{g_s}{N}\sum_{a,\k} \e^{\gamma\phi\eta}(-f'v^{\a\b}v^{\b\eta}v^\phi)_{a,\k}\frac{1}{4\gamma^3}
\nonumber \\
& & +  \Delta \s_{\a\b\b,\gamma},
\label{eqn:EMCHA} 
\end{eqnarray}
where $v^{\a}_{a,\k}\equiv\d_\a \e_{a,\k}$ and $v^{\a\b}_{a,\k}\equiv\d_\a \d_\b \e_{a,\k}$.
$\gamma=1/2\tau$ is the quasiparticle (QP) damping rate,
$f'$ is the $\e$-derivative of Fermi distribution function,
$g_s=2$ is the spin degeneracy, and $\e^{\a\b\g}$ is Levi-Civita symbol.
Here, $\Delta \s_{\a\b\b,\gamma}\equiv e^4\frac{g_s}{N}\sum_{a,\k} \e^{\gamma\phi\eta}(f'v^{\a\eta} v^{\b\phi}v^\b)_{a,\k}\frac{1}{8\gamma^3}$,
which vanishes when $\a=\b$.

In the present model with $\eta I_{\rm st}\ne0$, 
the symmetries of $v_\k^{\mu,\pm}$ and $v_\k^{\mu\nu,\pm}$ are summarized in Table \ref{tab:tab1}.
This table is useful to understand what kinds of nonreciprocal conductivities become finite.
Importantly, $v_{b,\k}^{\mu,-}$ is even in $\k$ and $v_{b,\k}^{\mu\nu,-}$ is odd in $\k$.
These facts give rise to the $B_z$-tunable nonreciprocal conductivities.
We show ${\hat v}_\k^{\Sigma}\equiv \sum_b v_{b,\k}^{\Sigma} (-f'(\e_{b,\k}))$ ($\Sigma=y,-$ or $zz,-$) in the $k_z=\pi/2$ plane in Fig. \ref{fig:fig3} (d) to clarify its functional form on the Fermi surface.
According to the discussion in Sect. \ref{sec:Symmetry} and TABLE \ref{tab:tab1},
the NLH is finite for $\s_{y\mu\mu}$ ($\mu=x,y,z$).
Also, the eMChA is finite for $\s_{z\mu\mu,x}$ and $\s_{x\mu\mu,z}$.

\begin{table} 
\caption{\label{tab:tab1}
(a). $\k$-dependences of $v^{\mu,\pm}$ and $\Omega^{\mu,\pm}$, and
(b). $\k$-dependences of $v^{\mu\nu,\pm}$ in the LC state with AB-stripe CDW.
Note that each function in the table should be multiplied by an $A_{1g}$ (totally symmetric) function in taking the derivative.
} 
\begin{ruledtabular} 
\begin{tabular}{|c||c|c|c|}
\hline
$\mu \ =$ & $x$ & $y$ & $z$ \\ 
\hline
$v^{\mu,-}$ & $k_x k_y$ & $1$ & $k_y  k_z$  \\
\hline
$v^{\mu,+}$ & $k_x$  & $k_y$  & $k_z$  \\
\hline \hline
$\Omega^{\mu,-}$ & $k_x k_z$ & $k_y k_z$ & $1$ \\
\hline
$\Omega^{\mu,+}$ & $k_x k_y k_z$ & $k_z$ & $k_y$ \\
\hline
\end{tabular} 
\vspace{3mm}
\begin{tabular}{|c||c|c|c|c|c|}
\hline
$\mu\nu \ =$ & $xx$, $yy$ & $zz$ & $xz$ & $yz$ & $xy$ \\ 
\hline
$v^{\mu\nu,-}$ & $k_y$ & $k_y$ & $k_x k_y k_z$ & $k_z$ & $k_x$ \\
\hline
$v^{\mu\nu,+}$ & $1$ & $1$ & $k_x k_z$ & $k_y k_z$ & $k_x k_y$ \\
\hline
\end{tabular} 
\end{ruledtabular} 
\end{table}

In the numerical study, we derive $v^{\a\b}_{a,\k}\equiv\d_\a \d_\b \e_{a,\k}$ from the sum rule $v^{\a\b}_{a,\k}={\tilde v}^{\a\b}_{a,\k}+2g^{\a\b}_{a,\k}$ to improve numerical accuracy.
Here, ${\tilde v}^{\a\b}_{a,\k}\equiv \langle u_{a,\k}|(\d_\a\d_\b {\hat H}^0_\k)| u_{a,\k} \rangle$, and $g^{\a\b}_{a,\k}$ is given by
\begin{eqnarray}
g^{\a\b}_a=\frac12 \sum_{b\ne a} \frac{{\rm Re}[v^{\a}_{ab}v^{\b}_{ba}+v^{\b}_{ab}v^{\a}_{ba}] \ \e_{ab}}{\e_{ab}^2+\delta^2},
\label{eqn:qqm}
\end{eqnarray}
where $\delta=0$ corresponds to the sum rule for the free-electron,
while we introduce $\delta=0.001$ to obtain reliable numerical results.
($\delta$ would correspond to Im$\Sigma^{\rm A}$.)
Note that Eq. (\ref{eqn:qqm}) is derived from the covariant differential of the operator $v^z=\d_z H^0$: $D_z v^z = \d_z v^z -i [A^z,v^z]$, where $A^z_{a,b}=i\langle u_a|\d_z u_b\rangle$ is the Berry connection.
We call $g^{\a\b}_a$ as the quasi-quantum metric (qQM) of the Bloch wavefunction because of its close relationship to the quantum metric (QM) given by
\begin{equation}
    \begin{split}
    G^{\a\b}_{a}
    &=\sum_{l\ne n} \frac{{\rm Re}[ v_{ab}^{\a} v_{ba}^{\b}+v_{ab}^{\b}v_{ba}^{\a} ]}
    {\epsilon_{ab}^2+\delta^2}
    \end{split},
\label{eqn:G-exp}
\end{equation}
where $g^{\a\b}_a$ and $G^{\a\b}_a$ differ in the power in $\epsilon_{ab}$ in the denominator.
In particular, both qQM and QM take huge values when the conduction and valence bands are nearly degenerated.
In later sections, we find that the giant eMChA conductivity in kagome metal is caused by the $\eta$-odd part of the QM ($G^{zz,-}$) at the Fermi level,
which gives a resonantly enhanced contribution for $\eta\lesssim0.02$.



Finally, we briefly discuss the reciprocal conductivities due to $\e_\k^-$ based on  Eqs. (\ref{eqn:NLH}) and (\ref{eqn:EMCHA}).
The NLH conductivity is given as
$\s_{yzz}\propto \langle v^{yzz} \rangle \gamma^{-2}$, where $v^{yzz}\equiv\d_y v^{zz}$,
by using the integration by parts.
Apparently, $\langle v^{yzz} \rangle$ is finite thanks to $\e_\k^-$. 
(Note that the Bloch's theory \cite{Bloch} does not prohibit a finite $\langle v^{yzz}\rangle$.)
Also, the eMChA conductivity is given as
$\s_{zzz,x} \propto \langle v^{yzz}v^{zz}-v^{zzz}v^{yz} \rangle \gamma^{-3}$,
whose $\eta$-odd term is finite according to Table \ref{tab:tab1}.
Thus, the TRS breaking dipole potential $\e_\k^{-}$ gives finite NLH and eMChA conductivities.

\subsection{Impact of QM on $\tau^3$-eMChA conductivity}
\label{sec:QM}

Here, we demonstrate the important role of the resonantly enhanced QM on the eMChA conductivity.
Here, we denote $\s_{zzz,x}\equiv \frac1N \sum_{\k}\s_{zzz,x}(\k)$,
where $\s_{zzz,x}(\k)\propto -\sum_a [f'((v^{zz})^2 v^y - v^{zz}v^{yz}v^z)]_a \frac1{\gamma^3}$
according to Eq. (\ref{eqn:EMCHA}).
$\eta$-even part $\s_{zzz,x}^+(\k)$ vanishes after the $\k$-summation, 
so only $\eta$-odd part $\s_{zzz,x}^-(\k)$ contributes to $\s_{zzz,x}$.
Figure \ref{fig:fig3} (e) shows $\s_{zzz,x}^{-}(\k)$ in the $k_z=\pi/2$ plane and the $k_x=0.08\pi$ plane, for $\eta=0.024$, $I_{\rm st}=0.01$, $T=0.01$.
It shows large, resonant-like peaks near the LC-induced band reconstruction between the pure-type and mix-type Fermi surfaces, denoted as red and blue broken lines in Fig. \ref{fig:fig3} (e).
(The original Fermi surfaces at $\eta=0$ are expressed in Extended Data Fig. \ref{fig:figS1} in the folded BZ.)
Therefore, giant eMChA conductivity in kagome metals originates from the ``pure-mix band reconstruction''; see Sect. \ref{sec:dxz-dyz}.


The resonant-like peaks in Fig. \ref{fig:fig3} (e) mainly originate from $[(v^{zz})^2]^-$ in $\s_{zzz,x}^{-}(\k)$.
Based on the ($a,b$)-band model,
an approximate relation $[(v^{zz})^2]^{-}_a \approx [(g^{zz})^2]^{-}_a\approx 2G^{zz,-}_a\cdot (v^z_{ab} v^z_{ba})^{+}$ is derived.
To verify this relation, we shows the $\eta$-odd part of the QM on the Fermi surface: $G^{zz,-}_\k\equiv \sum_a G^{zz,-}_{a,\k}(-f'(\e_{a,\k}^{0}))$.
Its peak positions are almost the same as those of $\s_{zzz,x}^{-}(\k)$ shown in Fig. \ref{fig:fig3} (e).
In fact, 
$\s_{xxx,z}^-(\k)\sim \sum_a f' G^{zz,-}_a v^{y,+}_a \frac1{\gamma^{3}}$
and $v^{y,+}_a$ is an odd-function of $k_y$.
Therefore, the giant eMChA in kagome metal originate from the $\eta$-odd QM $G^{zz,-}_\k$.
Three-dimensional mix-type Fermi surface with large $k_z$-directional velocity gives rise to large $G^{zz,-}_\k$.

\section{Numerical study of nonreciprocal conductivities}
 
Here, we perform the numerical study of the nonlinear conductivities,
under the stripe CDW along the AB direction (along the $y$-axis).
We put $\eta=0.008-0.024$, $I_{\rm st}=0.01$ and $T=0.01$ in the energy unit eV.
The obtained eMChA are almost unchanged for $T=0.005$. 
To achieve enough numerical accuracy,
we use $480\times480\times240$ $\k$-meshes in the original BZ.
The obtained eMChA is essentially unchanged even when 
$240\times240\times120$ $\k$-meshes are used.
We set $\gamma=1$, $e=1$ and $\hbar=1$ for simplicity.

\subsection{Giant NLH and eMChA}

\begin{figure*}[htb]
\includegraphics[width=.7\linewidth]{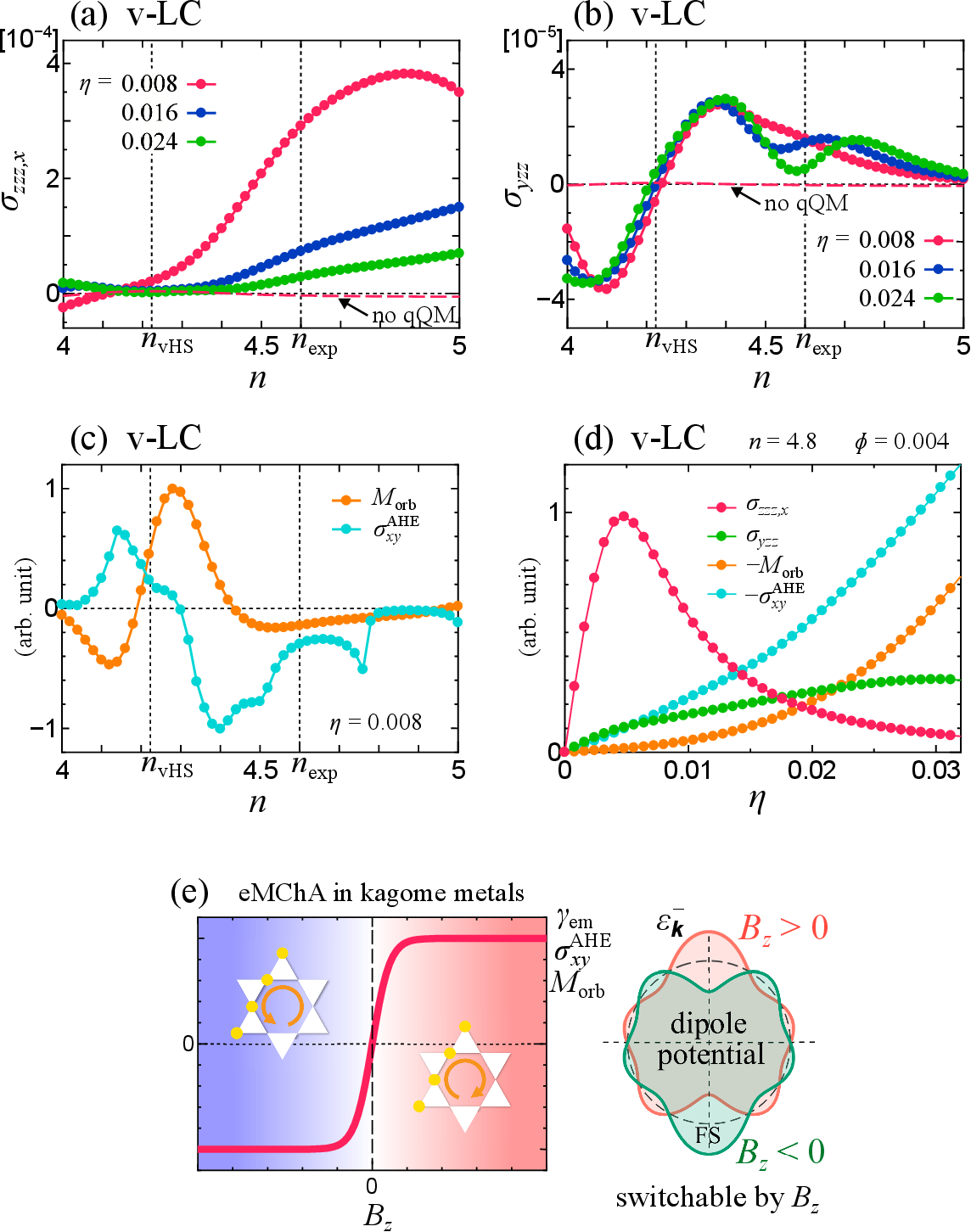}
\caption{
{Obtained NLH and eMChA conductivities}: \
(a) eMChA and (b) NLH conductivities in the v-LC state with AB direction $2a_0$ stripe CDW ($I_{\rm st}=0.01$) as functions of $n$ for $\eta=0.008-0.024$. 
We put $\gamma=1$ for simplicity. 
The numerical result is almost unchanged for the s-LC state; see Appendix D.
(c) $M_{\rm orb}$ and $\sigma_{xy}^{\rm AHE}$ as functions of $n$ in the v-LC state in the arbitrary unit. 
(d) $\eta$-dependences of the eMChA, NLH, $M_{\rm orb}$ and $\s_{xy}^{\rm AHE}$ as functions of $\eta$  at $n=4.8$, in the presence of small BO $\phi=0.004$. 
Only eMChA exhibits a resonance-like enhancement for $\eta\sim0.005$. 
(e) Jump in the eMChA, NLH, $M_{\rm orb}$ and AHE as function of $B_z$ below $\min\{T_{\rm LC},T_{\rm st}\}$, where the sign of $\eta$ is switched by the direction of $B_z$ while $|\eta|$ is fixed.
}
\label{fig:fig5}
\end{figure*}

Figures \ref{fig:fig5} (a) and (b) show the eMChA conductivity $\s_{zzz,x}$
and the NLH conductivity $\s_{yxx}$ in the v-LC state,
${\bm\eta}=\eta{\bm e}_{\rm I}$ and ${\bm\theta}=(0,0,0)$,
as functions of the electron filling $n$ at $T=0.01$.
(The obtained results are essentially unchanged for $T=0.001\sim0.01$.)
$n_{\rm exp}$ corresponds to the electron filling of CsV$_3$Sb$_5$,
and $n_{\rm vHS}$ is the vHS filling for the $d_{xz}$-orbital band.
Both $\s_{zzz,x}$ and $\s_{yxx}$ are odd-functions of $\eta$,
so they are reversed by changing the direction of $B_z$.
Unexpectedly, $\s_{zzz,x}$ increasing with decreasing the LC order 
for $\eta\ge0.008$, and $\s_{yxx}$ is nearly independent of $\eta$.
The same behaviors are obtained for $\eta=0.008\sim0.064$; see Appendix D.
Both $\s_{zzz,x}$ and $\s_{yxx}$ significantly decrease if $v_{b,\k}^{\mu\nu}$ in Eqs. (\ref{eqn:NLH}) and (\ref{eqn:EMCHA}) is replaced with ${\tilde v}_{b,\k}^{\mu\nu}$, that is, if the qQM $g_{b,\k}^{\mu\nu}$ is dropped.
Therefore, giant nonreciprocal conductivities originate from the qQM term, which is drastically enlarged near the band crossing points.

Considering that the LC transition temperature $T_{\rm LC}$ reported in kagome metals is $35\sim130$K, the magnitude of the order parameter at $T=0$ will be $|{\bm \eta}|= 0.005\sim0.02$ [eV] in the case of $|{\bm\eta}|/T_{\rm LC}\sim2$.
Therefore, giant eMChA observed in kagome metals is given by the resonantly magnified QM term.
In Appendix D, we find that 
$\s_{zzz,x}$ and $\s_{yxx}$ in the v-LC state in Fig. \ref{fig:fig5} (a) and (b) are almost unchanged in the s-LC state.
They are qualitatively unchanged if the $2\times2$ LC and BO coexist.

In Appendix E, we recover the unit of the conductivities given by the numerical study in Fig. \ref{fig:fig5} and Extended Data Fig. \ref{fig:figS4} (f).
We obtain $\rho_{zz}^{\rm exp}\approx 40000\gamma$ [$\mu\Omega$cm]
and $\rho_{xx}^{\rm exp}=2500\gamma \ [\mu\Omega{\rm cm}]$.
Considering the experimental values $\rho_{zz}\approx 30$ [$\mu\Omega$cm] \cite{eMChA}
and $\rho_{xx}=0.4 \ [\mu\Omega{\rm cm}]$ \cite{Roppongi},
we expect that $\gamma\lesssim 1\times 10^{-3}$ [eV] in high-quality kagome metals for $T\lesssim5$K.
By calculating $\gamma_{\rm eM}=-\s_{zzz,x}/(\s_{zz})^2$ numerically,
we obtain $|\gamma_{\rm eM}|\sim G/\gamma$, where $G\sim20\cdot(100I_{\rm st}{\rm [eV]})$ for $\eta\sim0.008$ and $n\approx n_{\rm exp}$.
Its unit is recovered as
$|\gamma_{\rm eM}^{\rm exp}|\sim10^{-17}\cdot(100I_{\rm st}/\gamma)$ [$A^{-1}T^{-1}m^2$].
Under finite $B_x$ measurement, the eMChA coefficient is magnified by the factor $z^2$, where $\displaystyle z\equiv \frac{\rho_{zz}(B_x)}{\rho_{zz}(0)}$.
In kagome metals, $z\sim20$ for $B_x=18$T \cite{eMChA}.

In addition, we discuss the eMChA coefficient by using the numerical result of $\s_{zzz,x}$ in Fig. \ref{fig:fig5} (a) and the experimental resistivity $\rho_{zz}(18{\rm T})\sim500$ [$\mu\Omega{\rm cm}$] \cite{eMChA}.
In Fig. \ref{fig:fig5} (a), $\s_{zzz,x}\sim 2\times10^{-4}$ for $\eta\sim0.008$ and $I_{\rm st}\sim0.01$ at $n\sim n_{\rm exp}$.
Then, the unit is recovered as $|\gamma_{\rm eM}^{\rm exp}|\sim 10^{-12}/(\gamma \ [{\rm eV}])^3$ [$A^{-1}T^{-1}m^2$].
Thus, experimental giant eMChA coefficient $\gamma_{\rm eM}^{\rm exp}\sim10^{-11}$ [$A^{-1}T^{-1}m^2$] can be understood by putting $\gamma\lesssim10^{-3}$ [eV] and $I_{\rm st}\sim0.01$ [eV] in the present theory.


We also calculate $M_{\rm orb}$ and the anomalous Hall conductivity $\s_{xy}^{\rm AHE}$ for $\eta=0.008$ as functions of $n$ in Fig. \ref{fig:fig5} (c).
(Here, the unit of the vertical axis is arbitrary.)
Both $M_{\rm orb}$ and $\s_{xy}^{\rm AHE}$ changes their signs for $\eta\rightarrow-\eta$.
These quantities take large values for $n\approx n_{\rm exp}$,
and their absolute values become maximum for $n\approx n_{\rm vHS}$,

\subsection{Impact of $\eta$-odd QM on eMChA}

Figure \ref{fig:fig5} (d) shows $\s_{zzz,x}$, $\s_{yzz}$, $M_{\rm orb}$, and $\s_{xy}^{\rm AHE}$ 
in an arbitrary unit as functions of $\eta$.
We set $n=4.8$ to avoid the exceptional $\eta$-dependence of $\s_{yxx}$ at $n=n_{\rm exp}$ shown in Fig. \ref{fig:fig5} (b).
Here, we introduce the small BO $\phi=0.004$ to make the numerical calculation stable for $\eta\sim0$. 
Importantly, only $\s_{zzz,x}$ exhibits the resonance-like singular behavior $\propto\eta^{-1}$ for $\eta\gtrsim0.005$.
This resonance-like behavior appears for a wide range of $n$.
(Although the maximum of $\s_{zzz,x}$ at $\eta\equiv \eta_{\rm max}$ decreases with $\phi$, the value of $\eta_{\rm max}\approx0.005$ is essentially independent of $\phi$; see Extended Data Fig. \ref{fig:figS4} (d).) 
In Fig. \ref{fig:fig5} (d), $M_{\rm orb}\propto \eta$ for $\eta\ll 0.01$ due to the $\eta$-$\phi$ bilinear term in $M_{\rm orb}$; see Appendix A.
However, $M_{\rm orb}\propto \eta^3$ is realized for $\eta\gtrsim 0.01$.

As we discussed in Sect. \ref{sec:QM}, 
$\s_{xxx,z}^-(\k)\sim \sum_a f' G^{zz,-}_a v^{y,+}_a \frac1{\gamma^{3}}$ 
is resonantly enhanced by $\eta$-odd QM term for small $\eta$.
In contrast, the NLH does not exhibit the resonance-like behavior
because the qQM $g^{\mu\nu}_{a,\k}$ in $\s_{yzz}(\k)$ drastically changes its sign around the Fermi surfaces as functions of $(a,\k)$, and its contributions are cencelled out upon the $\k$ integration.
In Appendix F, we demonstrate the relations $v^{zz,-}_\k\approx g^{zz,-}_\k$ and $((g^{zz})^2)^-_\k \propto G^{zz,-}_\k$.
It is found that $v^{zz,-}_\k$ exhibits the sudden sign changes near the resonant position.

Figure \ref{fig:fig5} (e) depicts the schematic picture of the present eMChA mechanism under the chiral LC state.
At a fixed $T (< T_{\rm LC})$, the LC order $\eta$ is fixed to $+\eta_0$ or $-\eta_0$, and
its sign is switched by $B_z$ because $M_{\rm orb}$ is the odd-function of $\eta$ in the triple-$\q$ LC state.
This fact leads to the sign reversal of the NLH and eMChA by tiny $B_z$ below $\min\{T_{\rm LC},T_{\rm st}\}$, where $T_{\rm st}$ is the stripe CDW transition temperature.
This result is consistent with the experimental report of the eMChA 
in Ref. \cite{eMChA}.
We stress that $\s_{zzz,x}$ is proportional to the AB direction order $\eta_1$, 
so the eMChA is induced by the single-$\q$ LC when its direction is parallel to the stripe CDW.
Importantly, in Ref. \cite{eMChA-K}, no eMChA is observed in KV$_3$Sb$_5$,
in which the triple-$\q$ LC emerges while no stripe CDW is observed experimentally.
This observation is naturally explained by the present scenario.

\section{Summary}

In this paper, we have identified the microscopic mechanism underlying the giant eMChA originating from chiral LC in kagome metals,
based on the derived exact formulas for eMChA and NLH with respect to the dominant terms for $\tau v_{\rm F}\gg1$.
Under the chiral LC order $\eta$, significant eMChA conductivity ($\propto \eta \tau^3 I_{\rm st}$) and NLH one ($\propto \eta \tau^2 I_{\rm st}$) manifest in the presence of $2a_0$ or $4a_0$ stripe CDW that is universally observed in CsV$_3$Sb$_5$.
The eMChA originates from the TRS breaking dipole potential $\e_\k^{-}\propto k_y\eta I_{\rm st}$.
Notably, eMChA experiences a dramatic amplification (by $\sim100$ times) due to resonantly enhanced QM effects arising from band reconstruction in the LC phase across a broad parameter range. 
This mechanism leads to substantial eMChA even when the LC order $\eta$ is small (approximately $\sim T_{\rm LC}=30\sim100{\rm K}$), underscoring the critical role of quantum geometry in kagome metals.
Importantly, in Ref. \cite{eMChA-K}, no eMChA is observed in KV$_3$Sb$_5$, in which the triple-$\q$ LC emerges while no stripe CDW is observed experimentally.
This observation is naturally explained by the present theory.
The present study manifests that kagome metal is an ideal platform for developing quantum geometric effects in strongly correlated metals.

The primary achievements of this study are summarized as follows:
(i) Derivation of essential electronic properties and their symmetries in multi-quantum orders of kagome metals through the experimental reports of the eMChA \cite{eMChA}.
(ii) Discovery of a novel QM mechanism responsible for giant and field-tunable eMChA, highlighting the importance of quantum geometry in kagome metals.
(iii) Development of nonreciprocal transport theory utilizing the QM in strongly correlated metals, like various kagome metals, chiral magnets \cite{CrNb3S6,MnSi}, and loop-current phase of cuprate superconductors \cite{Tazai-Morb}.

Here, we briefly discuss several experimental evidence of the LC state.
Recent transport measurement of highly symmetric fabricated CsV$_3$Sb$_5$ micro sample \cite{Moll-hz}
reveals that small magnetic field $h_z \ (<10{\rm T})$
or small strain strongly magnifies the LC order in the BO phase.
The realized LC+BO coexisting state becomes nematic as shown in Fig. 4 (c) of Ref. \cite{Tazai-kagome2} based on the GL free-energy theory.
The $h_z$-induced enhancement of the LC order is also observed by $\mu$SR measurements \cite{muSR4-Cs,muSR2-K,muSR5-Rb} and field-tuned chiral transport study \cite{eMChA}.
Also, the nematic LC+BO coexisting state has been reported by the STM measurement \cite{Madhavan}.
In addition, recent magnetic torque measurements \cite{Asaba} reveal the emergence of the single-$\q$ LC order above $T_{\rm BO}$, and the single-$\q$ to triple-$\q$ LC transition occurs under the conical magnetic field.
We stress that the eMChA is induced by the single-$\q$-LC when its direction is aligned with the stripe CDW in the present mechanism.
We note that the LC order will have a significant impact on the superconductivity, such as leading to the pair-density-wave and the $4e, 6e$ pairing states \cite{PDW-theory,Raghu-PDW,Wu-6e,Varma-6e,Varma-6e}.


We comment on a different kagome metal with double-layer structure ScV$_6$Sn$_6$, which exhibits the $\sqrt{3}\times\sqrt{3}$ CDW \cite{166-CDW}.
It was proposed that the CDW originates from the flat phonon modes with Sn vibrations \cite{Bernevig-166,Bernevig-166-2}.
Interestingly, ScV$_6$Sn$_6$ exhibits the spontaneous TRS breaking in the CDW phase \cite{166-mSR}.
The LC-induced QM effects in such different kagome metals would be an interesting future problem.

In this paper, we studied the eMChA and NLH conductivities in the LC phase due to the nonlinear Drude mechanism.
When the stripe CDW is along the $y$-direction (=AB direction), the $\eta$-odd qQM dipole $g^{zz,-}\propto k_y$ emerges.
Thus, the present nonlinear Drude mechanism gives
$\s_{zzz,x} \sim \tau^3\eta I_{\rm st}$ ($\s_{zzz,y}=0$),
which is proportional to $\tau^3$ and reversible under $B_z$.
The present eMChA mechanism is fundamentally different from the Berry-curvature dipole mechanism studied in Ref. \cite{eMChA} because the latter is proportional to $\tau^2$ and non-reversible under $B_z$; see the discussion in Appendix C.
The present mechanism is also different from the previously studied QM dipole mechanism that is proportional to $\tau^1$, which is considerably smaller than the present $\tau^3$-eMChA conductivity in kagome metals at low temperatures ($\tau v_{\rm F}\gg a_0$). 
Therefore, the present theory provides a convincing explanation for the switchable giant eMChA, which is a key experiment in the LC phase in kagome metals.


\acknowledgements
We are grateful to S. Onari, Y. Matsuda, T. Shibauchi, K. Hashimoto, and T. Asaba for fruitful discussions.
This study has been supported by Grants-in-Aid for Scientific Research from MEXT of Japan (JP24K00568, JP24K06938, 24H02231, 23K25816, 23K17665, JP22K14003), and JST CREST (Grant No. JPMJCR19T3).








\appendix

\section{GL free energy for kagome metals}
 
The cooperation and the competition between six LC and BO order parameters, 
${\bm \eta}$ and ${\bm \phi}$,
are well described by minimizing the Ginzburg-Landau (GL) free-energy.
In 2D kaogme lattice models,
the Ginzburg-Landau (GL) free energy without magnetic field 
(${\bm B}={\bm0}$) is $F=F_\eta+F_\phi+F_{\eta,\phi}$ 
\cite{Balents2021,Tazai-kagome2},
where
\begin{eqnarray}
&&\!\!\!\!\!\!\!\!\!
F_\phi= a_{\rm b} |{\bm \phi}|^2+ b_1\phi_1\phi_2\phi_3 
\nonumber\\
& &+d_1 (\phi_1^4+\phi_2^4+\phi_3^4)+ d_2 (\phi_1^2\phi_2^2+({\rm cycl.})) ,
\label{eqn:Fphi}
\\
&&\!\!\!\!\!\!\!\!\!
F_\eta= a_{\rm c} |{\bm \eta}|^2+ d_3 (\eta_1^4+\eta_2^4+\eta_3^4)+ d_4 (\eta_1^2\eta_2^2+({\rm cycl.})),
\label{eqn:Feta}
\end{eqnarray}
and $F_{\eta,\phi}$ contains the current-bond cross terms
proportional to $\eta^2\phi^1$ and $\eta^2\phi^2$:
\begin{eqnarray}
&&\!\!\!\!\!\!\!\!\!
F_{\eta,\phi}= b_2(\phi_1\eta_2\eta_3 + ({\rm cycl.}))
\nonumber\\
& &+2d_5(\phi_1^2\eta_1^2+\phi_2^2\eta_2^2+\phi_3^2\eta_3^2)
+ d_6(\phi_1^2\eta_2^2+({\rm cycl.})).
\label{eqn:Fphi-eta}
\end{eqnarray}
where the relation $b_1\sim -b_2$ is satisfied
\cite{Tazai-kagome2}.
Here, $a_x=r_x(T-T_x^0)$ ($x=$c or b),
where $T_{\rm c(b)}^0$ is the current-order (BO) 
transition temperature without other orders.
Theoretically,
$a_x\sim N(0)(-1+\lambda_x^{-1})$,
where $N(0)$ is the density-of-states ($\sim1 \ {\rm eV^{-1}}$)
and $\lambda_x$ is the DW equation eigenvalue 
of the current order ($x=$c) or the bond order ($x=$b) 
\cite{Tazai-Matsubara}.
$\lambda_x=1$ at $T=T_x^0$,
while $\lambda_x=0$ ({\it i.e.}, $a_x=\infty$)
in the absence of interaction.

This GL free energy theory can be extended to the 3D kagome metals.
Hereafter, we assume the $2\times2\times2$ order parameters.
When only the BO exists, 
we set ${\bm\phi}=(\phi_1,\phi_2,\phi_3)$ for the first layer, and 
${\bm\phi}'=(e^{i\psi_1}\phi_1,e^{i\psi_2}\phi_2,e^{i\psi_3}\phi_3)$ 
for the second layer.
Here, ${\bmpsi}=(\psi_1,\psi_2,\psi_3)$ is the set of the phase shifts.
The triple-$\q$ BO is stabilized by the 3rd-order $b_1$-term,
which is finite when $\psi_1+\psi_2+\psi_3=0$ (mod $2\pi$)
\cite{Tazai-Morb}.
By considering the 3rd-order GL free energy, we hereafter study the following two cases:
(i) ${\bmpsi}=(0,0,0)$, which represents the $2\times2\times1$ vertical-stack BO.
(ii) ${\bmpsi}=(0,\pi,\pi)$, $(\pi,0,\pi)$, and $(\pi,\pi,0)$,
each of which represents the $2\times2\times2$ shift-stack BO.

When only the LC exists, we set ${\bm\eta}=(\eta_1,\eta_2,\eta_3)$ for the first layer, and 
${\bm\eta}'=(e^{i\theta_1}\eta_1,e^{i\theta_2}\eta_2,e^{i\theta_3}\eta_3)$ 
for the second layer.
Here, ${\bm\theta}=(\theta_1,\theta_2,\theta_3)$ is the set of the phase shifts.
When ${\bm B}={\bm0}$, the 3rd-order term $b_1'\eta_1\eta_2\eta_3$ is absent in the GL free energy  ({i.e.}, $b_1'=0$) because it should be TRS invariant.
Then, the triple-$\q$ LC ${\bm\eta}_\a=\eta{\bm e}_\a$ ($\a=0$ or  $\a={\rm I}\sim{\rm III}$)
appears via the 2nd-order phase transition when $d_4/2d_3<1$.
In the opposite case, the single-$\q$ LC appears.

Now, we consider the LC+BO state with $|{\bm\phi}_0|\gg|\eta|$.
Then, we set the 3D BO ${\bm\phi}=\phi {\bm e}_0$, where ${\bm e}_0\equiv(1,1,1)/\sqrt{3}$.
This BO is realized when $b_1<0$.
We also introduce ${\bm e}_{\rm I}\equiv(1,-1,-1)/\sqrt{3}$, ${\bm e}_{\rm II}\equiv(-1,1,-1)/\sqrt{3}$, and ${\bm e}_{\rm c}\equiv(-1,-1,1)/\sqrt{3}$.
Then, the LC order parameter inside this 3D BO state
becomes ${\bm\eta}=\eta {\bm e}_\a$ $(\a={\rm I}\sim {\rm III})$ and ${\bmthe}={\bmpsi}$ 
to gain the $b_2$-term free energy for $b_2\sim -b_1>0$.
Thus, the LC+BO state exhibits the $Z_3$ nematic states,
and v-LC (s-LC) will appear inside the v-BO (s-BO) phase.
The 3D LC order for ${\bmthe}=(0,0,0)$ and that for 
${\bmthe}=(0,\pi,\pi)$ are shown in
Figs. \ref{fig:fig2} (e) and (g), respectively.



In the case of $T_{\rm BO}\gg T_{\rm LC}$, the BO is almost fixed as ${\bm\phi}_0=\phi{\bm e}_0$ for $b_1<0$.
For $T<T_{\rm LC}$, the LC order is obtained by minimizing the GL free energy
as ${\bar{\bm\eta}}_\a=\eta {\bar{\bm e}}_\a$, where ${\bar{\bm e}}_{\rm I}= (a,-1,-1)/A$, ${\bar{\bm e}}_{\rm II}= (-1,a,-1)/A$, and ${\bar{\bm e}}_{\rm III}= (-1,-1,a)/A$ with $A=\sqrt{2+a^2}$ 
\cite{Tazai-Morb}.
Here, we obtain $a\lesssim2$ for $T\lesssim T_{\rm LC}$ by minimizing the 3rd-order GL free energy, while $a$ approaches to $1$ as the ratio $|{\bm\eta}|/|{\bm\phi}|$ increases due to the 4th-order GL terms 
\cite{Tazai-Morb}.
The realized LC+BO order $({\bm\eta},{\bm\phi})=({\bar{\bm\eta}}_\a,{\bm\phi}_0)$ ($\a={\rm I}\sim{\rm III}$) is the $Z_3$ nematic state, which is observed experimentally 
\cite{Madhavan}.
In the main text, we set $a=1$ to simplify the discussion because the essential nature of the eMChA is unchanged.

Finally, we introduce the magnetic field $B_z$.
Then, the field-induced GL free energy is 
$\Delta F=-M_{\rm orb} B_z$, where 
the orbital magnetization along the $z$-axis is expressed as
\cite{Tazai-Morb}
%
\begin{eqnarray}
M_{\rm orb}&=& m_1 {\bm \phi}\cdot{\bm \eta} + m_2 \eta_1\eta_2\eta_3
\nonumber \\
& &+m_3(\eta_1\phi_2\phi_3 +\phi_1\eta_2\phi_3 +\phi_1\phi_2\eta_3)
\label{eqn:M-exp} .
\end{eqnarray}
Thus, the LC order ${\bm\eta}$ changes its sign depending on the direction of $B_z$.
This is important to understand the distinct field-induced change of the eMChA.
This result is naturally extended to the $2\times2\times2$ LC state.




\section{3D $d_{xz}$+$d_{yz}$ orbital kagome lattice model}

\begin{figure}[htb]
\includegraphics[width=.99\linewidth]{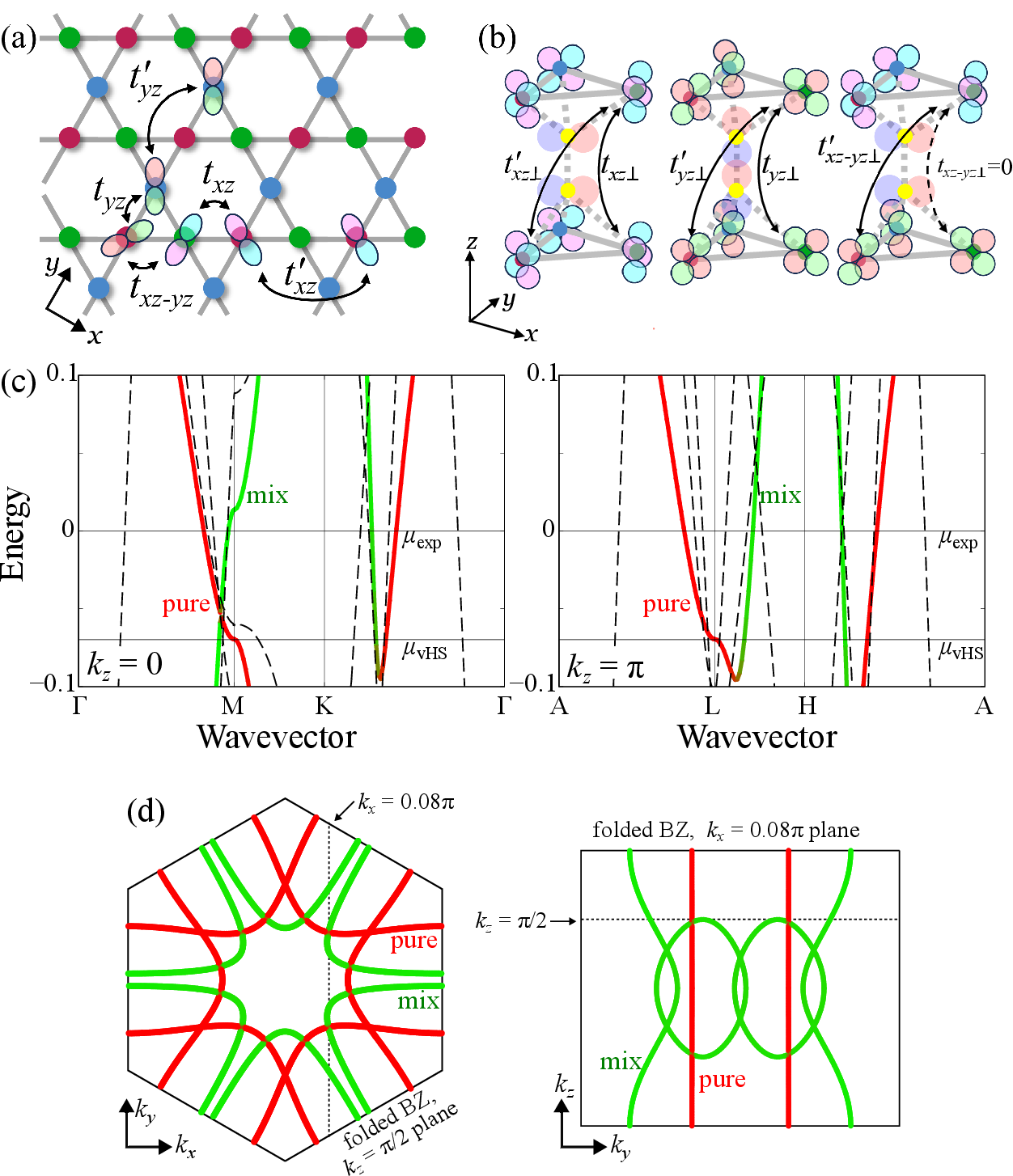}
\caption{
{3D kagome lattice model parameters}: \
(a) Intra-layer hopping integrals and (b) inter-layer hopping integrals in the present 3D kagome metal model.
In the present study, we drop small and unimportant hopping integrals $t_{yz}'$, $t_{{xz}\perp}$, $t_{{xz}\perp}'$, $t_{{xz-yz}\perp}$ and $t_{{xz-yz}\perp}'$ to simplify the model parameters. 
(c) Bandstructure of the present model in the $k_z=0$ plane and that in the $k_z=\pi$ plane. 
(d) Fermi surfaces in the folded BZ without the LC order ($\eta=0$).
When $\eta\ne0$, the Fermi surfaces are changed to those in Fig. 3 (e) in the main text.
}
\label{fig:figS1}
\end{figure}

Here, we introduce the 3D $d_{xz}$+$d_{yz}$ orbital kagome lattice model.
The intra-layer and inter-layer hopping integrals are shown in 
Extended Data Figs. \ref{fig:figS1} (a) and (b), respectively.
In Extended Data Fig. \ref{fig:figS1} (a),
the nearest and the third-nearest $d_{xz}$-$d_{xz}$
hopping integrals are $t_{xz}=-0.5$ and $t_{xz}'=-0.08$, respectively.
The nearest $d_{yz}$-$d_{yz}$ hopping integral is $t_{yz}=-1$, and the next-nearest one is $t_{yz}'=0$.
The nearest $d_{xz}$-$d_{yz}$ hopping integral is $t_{\rm xz-yz}=+0.05$ or $-0.05$, whose sign depends on the direction of the hopping by following the Slater-Koster theory.
In addition, we introduce the $d_{yz}$-orbital on-site energy $E_{yz}=+2.3$.

We also introduce the inter-layer hopping integrals shown in Extended Data Fig. \ref{fig:figS1} (b).
According to the first-principles study,
the $d_{yz}$-$d_{yz}$ inter-layer hopping $t_{{yz}\perp}$
(between A-A etc.) $t_{{yz}\perp}'$ (between A-B etc.) are much larger than other inter-layer hopping integrals.
Here, we set $t_{{yz}\perp}=t_{{yz}\perp}'=0.02$ to reproduce the $k_z$-dependence of the vHS energy of the mix-type band (=$d_{yz}$-orbital band)
\cite{ARPES-VHS,ARPES-band}.

At $n=n_{\rm exp} \ (=4.6)$
the present model reproduces the main $d$-orbital Fermi surfaces given by the first-principles study very well, as shown in Fig. 3 of the main text.
Extended Data Fig. \ref{fig:figS1} (c) shows the bandstructures in both the $k_z=0$ plane and the $k_z=\pi$ plane.
The characteristic bandstructure near the Fermi level,
such as the vHS saddle-point dispersion around M point,
are reproduced well.
Here, the $d_{xz}$ orbital weight is expressed as red color. 
Away from the Fermi level, the bandstructure starts to deviate from the DFT one.
This deviation gives no serious problem, because only the bandstructure near the Fermi level is important for the eMChA and the NLH conductivities, both of which are the QP transport phenomena ($\propto \tau^n$ with $n\ge1$).
Extended Data Fig. \ref{fig:figS1} (d) gives the Fermi surfaces in the folded BZ scheme. 
By introducing the LC order $\eta$, the Fermi surfaces are changed to those in Fig. 3 (e) in the main text.

In the main text, we perform the numerical study of the NLH and eMChA conductivites.
The obtained result will be quantitatively reliable for $n\approx n_{\rm exp}$, because the present model well reproduces the Fermi surfaces given by the first principles study.

\section{Derivations of linear and nonlinear conductivities and $M_{\rm orb}$}

Here, we derive the most divergent non-linear conductivites
with respect to the QP lifetime $\tau$,
which give the most significant contributions in good metals.
Here, $\gamma=1/2\tau$ is the QP damping rate
given as $-{\rm Im}\Sigma^R$.
In this section, we derive the NLH and eMChA conductivities
based on the Boltzmann equation method.

For this purpose, we introduce the orbital magnetic moment ${\bm m}_{n,\k}$,
the Berry curvature ${\bm \Omega}_{n,\k}$ of the $n$th band.
They are given as
\begin{equation}
    \begin{split}
    m^\mu_{n,\boldsymbol{k}}&=\frac{i}{2}\left[ \left\langle 
        \boldsymbol{\nabla}_{\boldsymbol{k}}u_{n,\boldsymbol{k}}\middle|
        \times\left(\epsilon_{n,\boldsymbol{k}}-\hat{h}(\boldsymbol{k})\right)\middle|
        \boldsymbol{\nabla}_{\boldsymbol{k}}u_{n,\boldsymbol{k}}\right\rangle \right]_\mu \\
        &=\frac{1}{2}\sum_{l\ne n} \frac{\e_{\mu\a\b} {\rm Im}[ v_{n,l}^{\a}v_{l,n}^{\b}] \ \epsilon_{nl,\boldsymbol{k}}}
        {(\epsilon_{nl,\boldsymbol{k}})^2+\delta^2}
    \end{split}
\label{eqn:m-exp}
\end{equation}
and
\begin{equation}
    \begin{split}
    \Omega^\mu_{n,\boldsymbol{k}}&=i\left[\left\langle\boldsymbol{\nabla}_{\boldsymbol{k}}
    u_{n,\boldsymbol{k}}\middle|\times\middle|\boldsymbol{\nabla}_{\boldsymbol{k}}
    u_{n,\boldsymbol{k}}\right\rangle\right]_\mu \\
    &=\sum_{l\ne n} \frac{\e_{\mu\a\b} {\rm Im}[ v_{n,l}^{\a}v_{l,n}^{\b} ]}
    {\left(\epsilon_{nl,\boldsymbol{k}}\right)^2+\delta^2} ,
    \end{split}
\label{eqn:Omega-exp} 
\end{equation}
where $\epsilon_{nl,\boldsymbol{k}}\equiv\epsilon_{n,\boldsymbol{k}}-\epsilon_{l,\boldsymbol{k}}$ and $\mu,\a,\b=x,y,z$.
Here, a small constant $\delta$ is introduced to obtain reliable numerical results, as we did for the QM and qQM in the main text. 
In deriving Eqs. (\ref{eqn:m-exp}) and (\ref{eqn:Omega-exp}),
we have used the relation
\begin{equation}
    \left\langle u_{l,\boldsymbol{k}}\d_{\mu} \middle|
    u_{m,\boldsymbol{k}}\right\rangle=\frac{v_{l,m}^{\mu}}
    {\epsilon_{l,\boldsymbol{k}}-\epsilon_{m,\boldsymbol{k}}}
\end{equation}
and $v_{l,m}^{\mu}=\left\langle
    u_{l,\boldsymbol{k}}\d^{\mu}\hat{h}(\boldsymbol{k})
    \middle|u_{m,\boldsymbol{k}}\right\rangle$ 
is the matrix element of the velocity operator 
for the wave vector $\boldsymbol{k}$ in the band representation.

Using both $\Omega^z$ and $m^z$,
the orbital magnetization $M_{\mathrm{orb}}$ along the $z$ axis is give as
\cite{Morb1S,Morb2S}
\begin{eqnarray}
M_{\mathrm{orb}}=\frac{1}{\pi N}\sum_{n,\boldsymbol{k}}
\left[m^z_{n,\boldsymbol{k}}f_{n,\boldsymbol{k}}
-\Omega^z_{n,\boldsymbol{k}}T\ln\left(1+e^{-(\epsilon_{n,\boldsymbol{k}}-\mu)/T}\right)\right].
\label{Morb}
\end{eqnarray}

Also, the anomalous Hall conductivity at $T=0$ is given as
\cite{Kontani-AHES}
\begin{widetext}
\begin{equation}
\sigma_{xy}^{\rm AHE}=\sigma_{xy}^{\mathrm{I}}+\sigma_{xy}^{\mathrm{IIa}}+\sigma_{xy}^{\mathrm{IIb}}.
\end{equation}
Here, $\sigma_{xy}^{\mathrm{I}}$, $\sigma_{xy}^{\mathrm{IIa}}$, 
and $\sigma_{xy}^{\mathrm{IIb}}$ are given by
\begin{equation}
    \begin{split}
    \sigma_{xy}^{\mathrm{I}}=&-\frac{1}{\pi N}
    \sum_{l\ne m,\boldsymbol{k}}\mathrm{Im}\left[v_{m,l,\boldsymbol{k}}^xv_{l,m,\boldsymbol{k}}^y\right]\mathrm{Im}\left[\frac{1}{\left(\epsilon_{l,\boldsymbol{k}}-\mu-i\gamma\right)
    \left(\epsilon_{m,\boldsymbol{k}}-\mu+i\gamma\right)}\right],
    \end{split}
\end{equation}
\begin{equation}
    \begin{split}
    \sigma_{xy}^{\mathrm{IIa}}=&-\frac{1}{\pi N}\sum_{l\ne m,\boldsymbol{k}}
    \frac{\mathrm{Im}\left[v_{m,l,\boldsymbol{k}}^xv_{l,m,\boldsymbol{k}}^y\right]}{\epsilon_{l,\boldsymbol{k}}-\epsilon_{m,\boldsymbol{k}}}
    \mathrm{Im}\left[\frac{\epsilon_{l,\boldsymbol{k}}+\epsilon_{m,\boldsymbol{k}}-2\mu-2i\gamma}{\left(\epsilon_{l,\boldsymbol{k}}-\mu-i\gamma\right)\left(\epsilon_{m,\boldsymbol{k}}-\mu-i\gamma\right)}\right],
    \end{split}
\end{equation}
and 
\begin{equation}
    \begin{split}
    \sigma_{xy}^{\mathrm{IIb}}=&\frac{2}{\pi N}
    \sum_{l\ne m,\boldsymbol{k}}
    \frac{\mathrm{Im}\left[v_{m,l,\boldsymbol{k}}^xv_{l,m,\boldsymbol{k}}^y\right]}{\left(\epsilon_{l,\boldsymbol{k}}-\epsilon_{m,\boldsymbol{k}}\right)^2}
    \mathrm{Im}\left[\ln\left(\frac{\epsilon_{l,\boldsymbol{k}}-\mu-i\gamma}{\epsilon_{m,\boldsymbol{k}}-\mu-i\gamma}\right)\right],
    \end{split}
\end{equation}
\end{widetext}
respectively.
Here, $\gamma \ (>0)$ is the imaginary part of the self-energy.
$\sigma_{xy}^{\mathrm{I}}$ and $\sigma_{xy}^{\mathrm{IIa}}$ are the contribution 
from the Fermi surface while $\sigma_{xy}^{\mathrm{IIb}}$ is the contribution 
from the Fermi sea by the Berry curvature: 
$\sigma_{xy}^{\mathrm{IIb}}\propto\sum_{n,\boldsymbol{k}}\Omega_{n,\boldsymbol{k}}f_{n,\boldsymbol{k}}$ 
for $\gamma\rightarrow0$ at $T\rightarrow0$.
We note that $\sigma_{xy}^{\rm AHE}$ also vanishes when the global TRS is preserved.

Next, we introduce the quasi-classical kinetic equations of the wave-packet 
\cite{Niu-revS}
\begin{eqnarray}
{\dot \r}&=& {\bm \nabla}_\k {\tilde \e}_\k -{\dot \k}\times{\bm \Omega}_\k,
 \\
{\dot \k}&=&-{\bm E}-{\dot \r}\times {\bm B},
\end{eqnarray}
where ${\tilde \e}_\k\equiv \e_\k - {\bm m}_\k\cdot{\bm B}$.
These coupled equations with respect to ${\dot \r}$ and ${\dot \k}$ are solved as
\begin{eqnarray}
D_\k{\dot \r}&=& 
{\bm \nabla}_\k {\tilde \e}_\k +{\bm E}\times{\bm\Omega}_\k+({\bm\nabla}_\k{\tilde \e}_\k\cdot{\bm\Omega}_\k){\bm B},
 \\
D_\k{\dot \k}&=&-{\bm E}-{\bm\nabla}_\k{\tilde \e}_\k\times{\bm B}-({\bm E}\cdot{\bm B}){\bm\Omega}_\k,
\end{eqnarray}
where $D_\k=1+{\bm B}\cdot{\bm\Omega}_\k$
\cite{Niu-revS}.
Then, the macroscopic current is given as
\begin{eqnarray}
&&{\bm j}=-\frac1N g_s \sum_\k D_\k {\dot \r}f,
\label{eqn:j-av}
\\
&&{\dot \k}\cdot {\bm \nabla}_\k f = \frac{f_0-f}{\tau} ,
\label{eqn:RTA}
\end{eqnarray}
where $f_0$ is the Fermi distribution function for $\e_\k$.
In the relaxation time approximation, the collision integral 
is expressed as $\frac{f_0-f}{\tau}$ in Eq. (\ref{eqn:RTA}),
where $\tau$ is the relaxation time of the wave-packet.
Then, we can expand the distribution function as $f=\sum_{n=0}^\infty f_n$, where $f_n\propto \tau^n$.
Then, Eq. (\ref{eqn:RTA}) is expressed as
\begin{eqnarray}
\sum_{n=0}^\infty {\dot \k}\cdot{\bm \nabla}_\k f_n = -\frac1\tau \sum_{n=1}^\infty f_n,
\end{eqnarray}
Thus, we obtain the recurrence formula $f_{n+1}=-\tau B f_n$,
where $\displaystyle B=\frac1{D_\k} \left( -{\bm E}-{\bm\nabla}_\k{\tilde \e}_\k\times{\bm B}-({\bm E}\cdot{\bm B}){\bm\Omega}_\k \right) \cdot {\bm \nabla}_\k$.
Using the derived $f_{n}$, we can derive the linear and nonlinear conductivities based on Eq. (\ref{eqn:j-av}).

First, we derive the linear conductivities in metals, by focusing the most divergent term with respect to $\tau=1/2\gamma$
\cite{Kontani-ROP}.
The longitudinal conductivity along the $\mu$-axis is given by the $E^1B^0$ term of $f_1$, where we are allowed to set ${\bm \Omega}_\k={\bm m}_\k={\bm0}$:
\begin{eqnarray}
\s_{\mu\mu}&=& -e^2 \frac{g_s}{N}\sum_{a,\k} f_a' (v^\mu v^\mu)_{a,\k}\frac1{2\gamma},
\label{eqn:sigma-2} 
\end{eqnarray}
where $a$ is the band index.
The Hall conductivity $\s_{\mu\nu}$ under the magnetic field $B_\a$ is given by the $E^1B^1$ term of $f_2$:
\begin{eqnarray}
\s_{\mu\nu}^{{\rm norm},\a}&=& B_\a e^3 \frac{g_s}{N}\sum_{a,\k} f_a' (v^\mu d^{\a,\nu})_{a,\k}\frac1{4\gamma^2},
\label{eqn:sigmaH} 
\end{eqnarray}
where $d^{\a,\nu}\equiv \e_{\a\b\delta}v^\b v^{\delta\nu}$,
and we set $B_\a=1$ at the end.
Note that $\s_{\mu\nu}^{{\rm norm},\a}$ is finite when $\mu \ne \nu \ne \a$.
Then, the Hall coefficient in the $\mu\nu$-plane is given by $R_{\rm H}=\s_{\mu\nu}^{{\rm norm},\a}/\s_{\mu\mu}\s_{\nu\nu}$.
Finally, the magnetoconductivity $\Delta\s_{\mu\mu}^\a$ under the magnetic field $B_\a$ is given by the $E^1B^2$ term of $f_3$:
\begin{eqnarray}
\Delta\s_{\mu\mu}^\a&=&-B_\a^2 e^4\frac{g_s}{N}\sum_{a,\k} f_a' ( \{ d^{\a,\mu}\}^2)_{a,\k}\frac1{8\gamma^3},
\label{eqn:Dsigma} 
\end{eqnarray}
which is always negative, and we set $B_\a=1$ at the end.
Then, the magnetoresistance (MR) under the magnetic field along the $\a$-axis up to $O(B_\a^2)$ is given by
$\rho_{\mu\mu}(B_\a)=\rho_{\mu\mu}(0)+[-\Delta\s_{\mu\mu}^\a/\s_{\mu\mu}-(\s_{\mu\nu}^{{\rm norm},\a})^2/\s_{\mu\mu}\s_{\nu\nu}]B_\a^2$,
where $\nu \ne \mu \ne \a$.

Here, we consider the MR along the $z$ axis $\rho_{zz}(B_x)$, which is $\eta$-even.
According to Eq. (\ref{eqn:Dsigma}), 
$\Delta\s_{zz}^\a\propto \sum_{a,\k} f_a' ( \{ d^{x,z}\}^2)_{a,\k}\gamma^{-3}$ and $d^{x,z}=v^y v^{zz}-v^z v^{yz}$.
The $(v^{\mu\nu})^2$-term enhances the $\rho_{zz}$ in the $2\times2$ LC state, due to the qQM term in $v^{\mu\nu}$.
We present the numerical analysis in Appendix E.

Next, we derive the most divergent nonlinear conductivities in noncentrosymmetric metals 
with respect to $\tau=1/2\gamma$.
The NLH and eMChA conductivities are given by the $E^2B^0$ term in $f_2\propto \tau^2$ and 
the  $E^2B^1$ terms in $f_3\propto\tau^3$, respectively.
They are derived from $f_i$ by setting ${\bm\Omega}_\k={\bm m}_\k={\bm0}$.
These most-divergent terms are finite when both IS and TRS are broken.
Here, we derive only $\s_{yzz}$ and $\s_{zzz,x}$ for simplicity.
Their expressions are obtained as
\begin{widetext}
\begin{eqnarray}
\s_{\a\b\b}&=& -e^3 \frac{g_s}{N}\sum_{a,\k} (\v^\a \d_\b \d_\b f)_{a,\k}\frac1{4\gamma^2},
\nonumber \\
&=& e^3 \frac{g_s}{N}\sum_{a,\k} f_a' (v^{\a\b}v^\b)_{a,\k}\frac1{4\gamma^2},
\label{eqn:NLH-2} \\
\s_{\a\b\b,\g}
&=& e^4g_s\sum_{a,\k} \e^{\gamma\phi\eta}
(v^\a (v^\phi \d_\eta)\d_\b \d_\b f + v^\a \d_\b (v^\phi \d_\eta)\d_\b f)_{a,\k}\tau^3
\nonumber \\
&=& e^4g_s\sum_{a,\k} \e^{\gamma\phi\eta}
(2 v^{\a\b\eta} v^\phi v^\b f' + v^{\a\eta} v^{\b\phi} v^\b f')_{a,\k}\tau^3 ,
\label{eqn:EMCHA-2} 
\end{eqnarray}
\end{widetext}
which are canceled the nonlinear Drude mechanism, and we set $B_\a=1$ in Eq. (\ref{eqn:EMCHA-2}).
(The relations $\e^{\gamma\phi\eta} v^\phi v^\eta = \e^{\gamma\phi\eta} v^{\phi\eta} =0$
are used in the derivation.)
These nonlinear conductivities are odd-functions of the LC order $\eta$.
The expression of $\s_{\a\b\b,\g}$ in the main text is given by the partial integral of the first term in the last line of Eq. (\ref{eqn:EMCHA-2}) with respect to $k_\eta$.
The second term in the last line of Eq. (\ref{eqn:EMCHA-2}) vanishes when $\a=\b$.

\begin{figure}[htb]
\includegraphics[width=.99\linewidth]{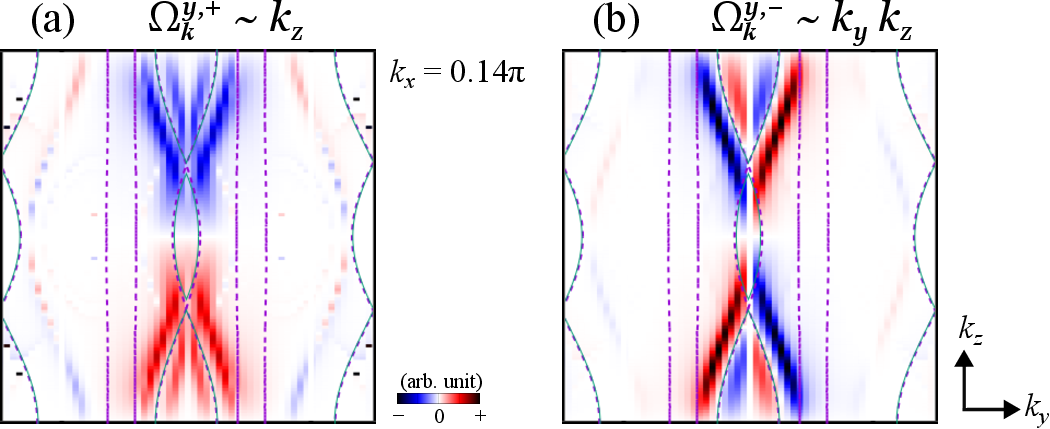}
\caption{
{Berry curvature of 3D kagome lattice model with LC order + stripe CDW}: \
Obtained (a) $\Omega_{\k}^{y,+}$ and (b) $\Omega_{\k}^{y,-}$ in the LC+stripe CDW state in the $k_x=0.14\pi$ plane.
It is found that $\Omega_{\k}^{y,+}\sim k_z$ and $\Omega_{\k}^{y,-}\sim k_y k_z$.
The dipole component $\Omega_{\k}^{y,+}\sim k_z$ gives rise to the NLH and eMChA, while their signs are unchanged by $\eta\rightarrow-\eta$.
}
\label{fig:figS3}
\end{figure}

In the presence of TRS, the nonlinear Drude terms in Eqs. (\ref{eqn:NLH-2}) and (\ref{eqn:EMCHA-2}) vanish.
In this case, the most divergent NLH and eMChA conductivities are given by the $E^2B^0$ term in $f_1\propto \tau^1$ and 
the  $E^2B^1$ terms in $f_2\propto\tau^2$, respectively.
They are expressed as the summation of the ${\bm\Omega}_\k$-linear and the ${\bm m}_\k$-linear terms.
Here, we derive only $\s_{\mu zz}$ and $\s_{zzz,\mu}$ ($\mu=x,y$) for simplicity.
Their expressions are obtained as
\begin{eqnarray}
\s_{\mu zz}&=& -e^2 \frac{g_s}{N}\sum_{a,\k} \e_{\mu\delta z} f_a' (v^z \Omega^{\delta})_{a,\k}\frac1{2\gamma},
\label{eqn:NLH-3} \\
\s_{zzz,\mu}
&=& -B_y e^3 \frac{g_s}{N}\sum_{a,\k} f_a'(v^{zz}\d_z m^{\mu}+v^z \d_{zz}m^{\mu})_{a,\k}\frac1{4\gamma^2},
\nonumber \\
& & -B_y e^3 \frac{g_s}{N}\sum_{a,\k} f_a'(2v^{zz}v^z \Omega^{\mu}+(v^z)^2 \d_{z}\Omega^{\mu})_{a,\k}\frac1{4\gamma^2},
\nonumber \\
\label{eqn:EMCHA-3} 
\end{eqnarray}
where we set $B_\a=1$ in Eq. (\ref{eqn:EMCHA-3}).
Here, we introduce $\Omega^{\mu,\pm}=[\Omega^{\mu}(\eta)\pm\Omega^{\mu}(-\eta)]/2$.
Extended Data Fig. \ref{fig:figS3} shows the obtained
(a) $\Omega_{\k}^{y,+}$ and (b) $\Omega_{\k}^{y,-}$ in the LC+stripe CDW state in the $k_x=0.14\pi$ plane.
In this figure, we show $\Omega^{y,\pm}_\k=\sum_a \Omega^{y,\pm}_{a,\k}(-f'(\e_{a,\k}^{0}))$ to clarify its functional form on the Fermi surface.
It is found that $\Omega_{k}^{y,+}\sim k_z$ and $\Omega_{\k}^{y,-}\sim k_y k_z$.
The symmetry properties of $\Omega_{\k}^{\mu,\pm}$ are summarized in Table I of the main text.
The symmetry of $m_{\k}^{\mu,\pm}$ is the same as that of $\Omega_{\k}^{\mu,\pm}$.
Thus, the dipole components $\Omega_{\k}^{y,+},m_{\k}^{y,+}\sim k_z$ give rise to the NLH $\s_{xzz}$ and eMChA $\s_{zzz,y}$.
However, neither $\Omega_{\k}^{y,-}$ nor $m_{\k}^{y,-}$ contribute to the eMChA and the NLH.
Therefore, Eqs. (\ref{eqn:NLH-3}) and (\ref{eqn:EMCHA-3})
are NOT switchable by the direction of $B_z$.
In addition, Eq. (\ref{eqn:EMCHA-2}) would dominate over Eq. (\ref{eqn:EMCHA-3}) in good metals with $\tau v_{\rm F} \gg 1$.
For these reasons, we study the nonlinear Drude mechanism
given in Eqs. (\ref{eqn:NLH-2}) and (\ref{eqn:EMCHA-2}) in the present paper.

\section{Numerical study of NLH and eMChA in LC+BO states}

\begin{figure}[htb]
\includegraphics[width=.99\linewidth]{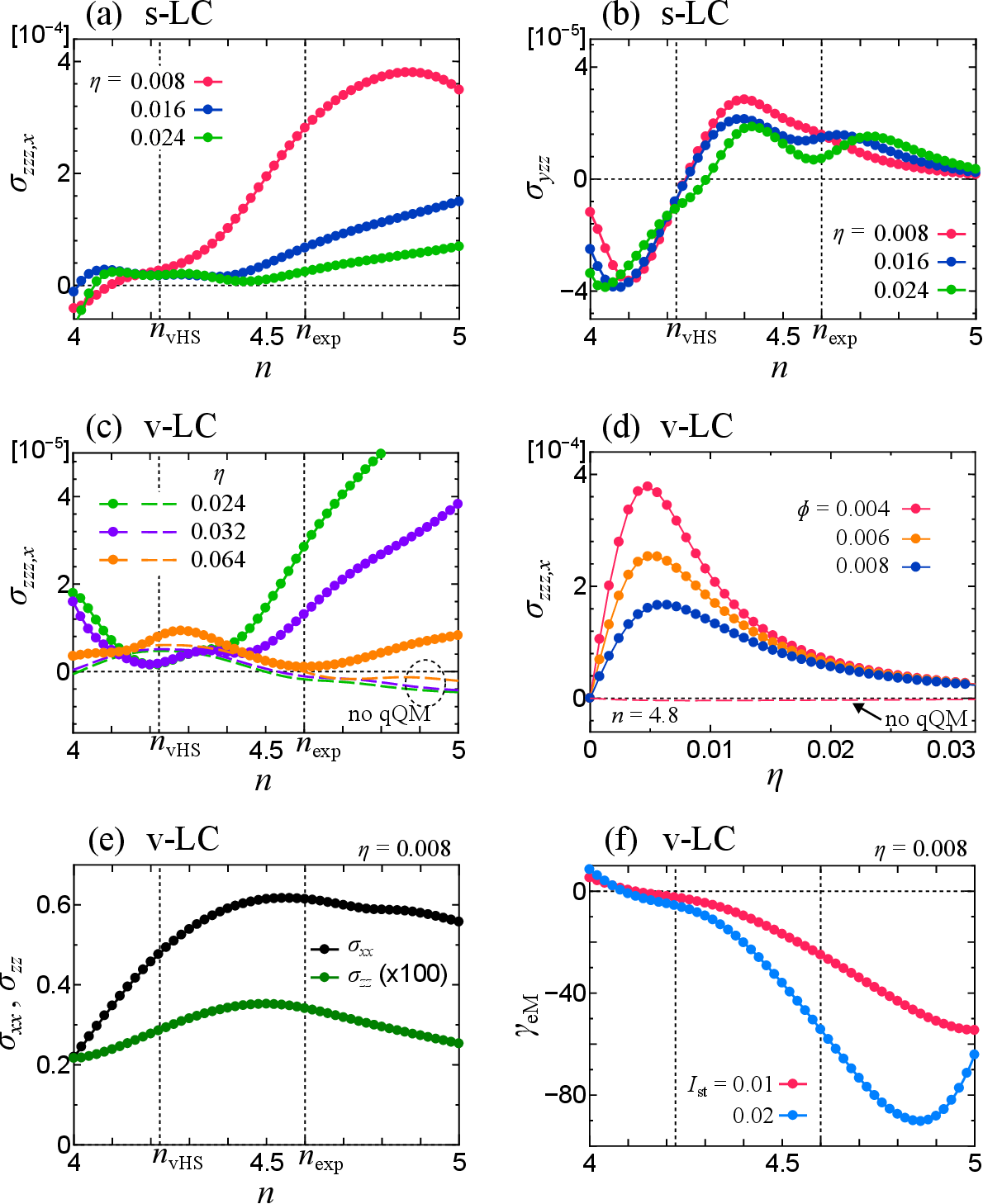}
\caption{
{NLH and eMChA conductivities in LC + BO state}: \
(a) eMChA and (b) NLH conductivities in the s-LC state with AB direction $2a_0$ stripe CDW ($I_{\rm st}=0.01$) as functions of the filling $n$.
The obtained results are very similar to those for the v-LC state shown in Fig. 4 in the main text.
We put $\gamma=1$ for simplicity. 
(c) eMChA in the v-LC state with $I_{\rm st}=0.01$ for larger $\eta$ ($0.024$, $0.032$ and $0.064$) as functions of $n$.
In this region, the eMChA monotonically decreases with $\eta$ and approaches to the non-resonant value without qQM in $v^{\mu\nu}$. 
(d) eMChA $\s_{zzz,x}$ as function of $\eta$ in the presence of the OB at $\phi=0.004$ and $0.008$. 
$\s_{zzz,x}$ is still large even for $\phi=0.008$.  
(e) $\s_{xx}$ and $\s_{zz}$ as functions of $n$ for $\eta=0.008$.
Here, we set $\gamma=1$. 
(f) $\gamma_{\rm eM}=-\s_{zzz,x}/(\s_{zz})^2$ for $I_{\rm st}=0.01$ and $0.02$.
Here, we set $\gamma=1$. 
}
\label{fig:figS4}
\end{figure}

In Fig. 4 of the main text, we calculated the NLH and eMChA due to the v-LC order and the stripe CDW. 
Here, we perform the same analysis for the s-LC state.
Extended Data Fig. \ref{fig:figS4} (a) and (b) show the obtained
$\s_{zzz,x}$ and $\s_{yzz}$ for the s-LC state, where
${\bm\phi}=\phi{\bm e}_0$, ${\bm\eta}=\eta{\bm e}_{\rm I}$, 
and ${\bm\psi}={\bm\theta}=(0,\pi,\pi)$ under the AB direction stripe CDW.
Here, we set $\eta=0.008\sim0.024$ and $I_{\rm st}=0.01$.
The obtained results are very similar to those for the v-LC state shown in Figs. 4 (a) and (b) in the main text. 
Therefore, the inter-plane stacking pattern is not essential for both eMChA and NLH conductivities.

In Fig. 4 (a) of the main text, 
we show the eMChA in the v-LC state for $\eta=0.008\sim0.024$, where the eMChA takes huge value due to the resonantly magnified QM.
In Extended Data Fig. \ref{fig:figS4} (c), we show the eMChA for larger $\eta$ ($0.024$, $0.032$ and $0.064$) at $I_{\rm st}=0.01$ as functions of $n$.
In this region, the eMChA monotonically decreases with $\eta$ and approaches to the non-resonant value without qQM in $v^{\mu\nu}$.
Considering that $T_{\rm LC}$ reported in kagome metals is $35\sim100$K, the magnitude of the order parameter at $T=0$ will be $|{\bm \eta}|= 0.005\sim0.015$ in the case of $|{\bm\eta}|/T_{\rm LC}\sim2$.
Therefore, giant eMChA observed in kagome metals is given by the resonantly magnified QM term, which is shown in Fig. 4 (a) of the main text.

We also study the effect of the triple-$\q$ BO on the eMChA, which becomes large inside the star-of-David BO state in kagome metals. 
Extended Data Fig. \ref{fig:figS4} (d) shows $\s_{zzz,x}$ obtained for the v-LC state
(${\bm\eta}=\eta{\bm e}_{\rm I}$ and ${\bm\psi}=(0,0,0)$)
and the v-BO state
(${\bm\phi}=\phi{\bm e}_0$ and ${\bm\theta}=(0,0,0)$).
This LC+BO order led to the $Z_3$ nematic state whose director is along the AB direction.
Here, we set $\eta=0\sim0.024$, $I_{\rm st}=0.01$, and $\phi=0.004\sim0.008$.
It is found that $\s_{zzz,x}$ is suppressed by $\phi$,
while it is still large even for $\phi=0.008$.

In Extended Data Fig. \ref{fig:figS4} (e),
we show the longitudinal linear conductivities
$\s_{xx}$ and $\s_{zz}$ as functions of $n$ for $\gamma=1$.
Here, we set $\eta=0.008$ and $\phi=0$, while the obtained results are nearly constant for $\eta,\phi\lesssim0.01$.
Both $\s_{xx}$ and $\s_{zz}$ are nearly $n$-independent.
These results lead to 
$\rho_{zz}=40000\gamma \ [\mu\Omega{\rm cm}]$ and
$\rho_{xx}=2500\gamma \ [\mu\Omega{\rm cm}]$;
see the discussion about the experimental units in Appendix E.

Extended Data Fig. \ref{fig:figS4} (f) exhibits 
$\gamma_{\rm eM}=-\s_{zzz,x}/(\s_{zz})^2$ for $I_{\rm st}=0.01$ and $0.02$.
Thus, $\gamma_{\rm eM}$ increases in proportion to $I_{\rm st}$ at $n\approx n_{\rm exp}$.

\section{Experimental units of $\s_{xxx,z}$ and $\gamma_{\rm eM}$}

In the main text, we performed the numerical study in the following units: $\hbar=e=1$, the unit of energy eV, the intra-layer nearest-neighbor distance $b_0\equiv |{\bm a}_{\rm AB}|=1$, and the inter-layer distance $c_0=1$.
In kagome metals, $b_0\approx0.3$nm and $c_0\approx1$nm.
Here, we restore the units of the conductivities in the 3D kagome lattice model with 3-site unit cell, where the volume of the unit cell is $v_{\rm uc}=2\sqrt{3}b_0^2 c_0$.
(In the 24-site unit cell model with the $2\times2\times2$ LC order, the unit cell volume becomes $8v_{\rm uc}$. Thus, the derived conductivities should be divided by 8 to obtain the values of the original 3-site unit cell model.)

First, we consider the expression of $\s_{zz}$ given in Eq. (\ref{eqn:sigma-2})
in Appendix.
To restore $\hbar$, $b_0$ and $c_0$, 
we magnify $\gamma$, $N$, $v^{x(y)}$, and $v^z$ with 
$\hbar^{-1}$, $v_{\rm uc}$, $b_0\hbar^{-1}$, and $c_0\hbar^{-1}$, respectively.
Then, we obtain
$\displaystyle \s_{zz}^{\rm exp}=\left(\frac{e^2}{\hbar}\right) \left(\frac{c_0}{2\sqrt{3}b_0^2}\right) \s_{zz}$,
where $\s_{zz} \propto 1/(\gamma \ [{\rm eV}])$ is the numerical result given in Eq. (\ref{eqn:sigma-2}).
Here, $\hbar/e^2 =0.42\times10^{4} \ [\Omega]$ and $2\sqrt{3} b_0^2/c_0=3.1\times10^{-10} \ [{\rm m}]$.
Thus, we obtain $\rho_{zz}=40000\gamma \ [\mu\Omega{\rm cm}]$
because $\s_{zz}\approx 0.0038/(\gamma \ [{\rm eV}])$ in the Extended Data Fig. \ref{fig:figS4} (e).
In the same way,
$\rho_{xx}=2500\gamma \ [\mu\Omega{\rm cm}]$
because $\s_{xx}\approx 0.6/(\gamma \ [{\rm eV}])$ in the Extended Data Fig. \ref{fig:figS4} (e).
Therefore, we obtain $\rho_{zz}=40\ [\mu\Omega{\rm cm}]$
and $\rho_{xx}=2.5\ [\mu\Omega{\rm cm}]$ for $\gamma=10^{-3}$ [eV].
Experimentally, $\rho_{zz}^{\rm exp}\approx30$ $\mu\Omega{\rm cm}$ at ${\bm B}=0$
\cite{eMChA},
and $\rho_{xx}=0.4 \ [\mu\Omega{\rm cm}]$ in high-quality samples 
\cite{Roppongi}.
Therefore, $\gamma\lesssim10^{-3}$ [eV] is expected to be realized at $T\approx0$.
The observed giant MR $\Delta\rho_{zz}(|{\bm B}|)$ indicates that 
the quantum limit $\omega_H \tau\gg1$ is satisfied for $|{\bm B}|\gtrsim5$ T
\cite{eMChA}.

Now, we consider $\gamma_{\rm eM}=-\s_{zzz,x}(\rho_{zz})^2$,
by which the nonlinear resistance is expressed as 
$\rho_{zz}=\rho_{zz}^0(1+\gamma_{\rm eM}B_x J_z)$.
(Exactly speaking, $\gamma_{\rm eM}^{abc,d}=-\rho_{aa'}(\s_{a'bb',d})\rho_{b'c}$, which gives the nonlinear resistance $\rho_{ae}({\bm B},{\bm j})=(\delta_{ab}+\gamma_{\rm eM}^{abc,d}j_cB_d)\rho_{bc}$.
Note that $\rho_{ac}\sigma_{cb}=\delta_{ab}$ \cite{Liu-eMChAS}.)
Based on the expression of $\s_{zzz,x}$ given in Eq. (\ref{eqn:EMCHA-2}) in Appendix, we can restore $\hbar$, $b_0$ and $c_0$ as explained above.
(Note that $v^{zz}$ and $v^{yz}$ are multiplied with $c_0^2/\hbar$ and $b_0 c_0/\hbar$, respectively.)
Then, we obtain
$\displaystyle \s_{zzz,x}^{\rm exp}=\left(\frac{e^2}{\hbar}\right)^2 \left(\frac{c_0^3}{2\sqrt{3}b_0^3}\right)\mu_B \left(\frac{\hbar}{e}\right) \left(\frac{2}{E_{b_0}}\right) \s_{zzz,x}$,
where $\mu_B= e\hbar/2m_e=5.8\times10^{-5} \ [{\rm eV} {\rm T}^{-1}]$ is Bohr magneton, where $m_e$ is the bare electron mass. 
$e^2/\hbar=2.4\times10^{-4} [\Omega^{-1}]$ and $\hbar/e=3.5\times10^{3} \ [{\rm eV}A^{-1}]$.
$E_{b_0}\equiv \hbar^2/m_e b_0^2$ is $0.89 \ [{\rm eV}]$ for $b_0=0.3$nm.
(Note that $E_{b_0}=m_e c^2\times10^{-6} = 0.51 \ [{\rm eV}]$ when $b_0=1000\lambdabar_e=3.9\times10^{-10} \ [{\rm m}]$, where $\lambdabar_e=\hbar/m_e c$; see Ref. \cite{Tazai-Morb}.)
The numerical study in Fig. \ref{fig:fig5} (a) gives
$\s_{zzz,x} =S_{\rm eM}/(\gamma \ [{\rm eV}])^3$, where
$S_{\rm eM}\sim 2\times10^{-4}$ for $\eta\sim0.008$ and $I_{\rm st}\sim0.01$ at $n\sim n_{\rm exp}$.
(The obtained $\s_{zzz,x}$ would not be affected by the $p$-orbital bands, which are dropped in the present model, considering the smallness of the $p$-$d$ hybridization.)
Then, we obtain $\gamma_{\rm eM}^{\rm exp}\sim 10^{-21}/(\gamma \ [{\rm eV}])^3$ [$A^{-1}T^{-1}m^2$]
if the experimental resistivity $\rho_{zz}(B_x)\sim500$ [$\mu\Omega{\rm cm}$] for $B_x=18$ T is applied.
When $\gamma\sim0.5\times10^{-3}$ [eV], 
the obtained $\gamma_{\rm eM}^{\rm exp}$ reaches $\sim10^{-11} \ [A^{-1}T^{-1}m^2]$.

In Ref. \cite{eMChA}, 
giant $\gamma_{\rm eM}\sim 10^{-11}$ $[{\rm A^{-1} T^{-1}m^2}]$ is reported in CsV$_3$Sb$_5$ by taking the difference between $B_x=18\ {\rm T}$ and $-18 \ {\rm T}$.
Although $\rho_{zz}(B_x)\approx30$ $\mu\Omega{\rm cm}$ at $B_x=0$.
$\rho_{zz}$ is drastically enlarged due to its large MR:
$\displaystyle z\equiv \frac{\rho_{zz}(18{\rm T})}{\rho_{zz}(0)}\approx20$ at $T=5$K.
The observed large MR does not saturate in the quantum limit $\omega_H \tau\gg1$ ($\omega_H\equiv e|{\bm B}|/m_H c$) due to the open-orbit Fermi surface along the $k_z$ axis.

We also study the expression $\gamma_{\rm eM}=-$[Eq.(\ref{eqn:EMCHA-3})]/[Eq.(\ref{eqn:sigma-2})]$^2$ for $\mu=z$.
Its experimental units are restored as
$\displaystyle \gamma_{\rm eM}^{\rm exp}\approx -\mu_B\left(\frac{\hbar}{e}\right)\frac{4\sqrt{3}b_0 c_0}{E_{b_0}}\gamma_{\rm eM} \approx -\frac{0.5\times10^{-18}}{E_{b_0}} \gamma_{\rm eM}$ $[{\rm A^{-1}T^{-1}m^2}]$,
where $\gamma_{\rm eM}$ is the present numerical result in unit ${\rm eV}^{-1}$.
Because $|\gamma_{\rm eM}| \sim 20/(\gamma \ [{\rm eV}])$ in the Extended Data Fig. \ref{fig:figS4} (f),
we obtain $|\gamma_{\rm eM}^{\rm exp}|\sim 10^{-14}\ [{\rm A^{-1}T^{-1} m^2}]$
for $\gamma\sim10^{-3}$.
It is magnified by the factor $z^2\gtrsim 10^2$
by taking account of the magnetoresistance.
Thus, the present theory can explain giant $\gamma_{\rm eM}$ in kagome metals by considering the giant enhancement factor $z^2$.

To summarize, the giant eMChA in kagome metal is given by the ``QM mechanism'' in the LC phase.
The present eMChA is drastically magnified by the ``pure-mix resonance'' due to the band-folding in the $2\times2$ LC phase.
In fact, the eMChA is proportional to $\sum_\k v^{y,+}((g^{zz})^2)^{-} f_{0}'\propto \sum_\k v^{y,+} G^{zz,-}f'$, the eMChA grows resonating way for $\eta\lesssim0.01$ due to the $G_a^{zz,-}$.
Interestingly, this resonance mechanism is not important for $M_{\rm orb}$, $\s_{yzz}$, and the anomalous Hall conductivity $\s_{xy}^{\rm AHE}$.
Thus, the ``QM-induced resonating enhancement'' is specific for the eMChA.

\begin{figure}[htb]
\includegraphics[width=.6\linewidth]{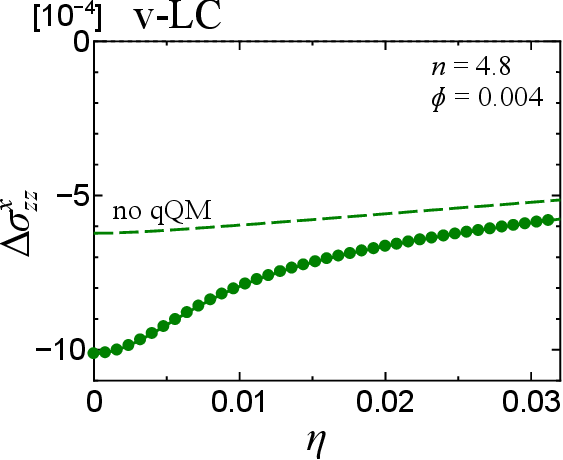}
\caption{
{Enhancement of magnetoconductivity by QM}: \
$\Delta \s_{zz}^x$ as a function of $\eta$. (Here, $\gamma=1$.)
$\Delta \s_{zz}^x$ is enlarged for $\eta\lesssim0.01$ by the $\eta$-even QM.
However, $\Delta \s_{zz}^x$ is already large without the QM,
so the QM-induced enhancement is only quantitative.
}
\label{fig:figS5}
\end{figure}

Finally, we briefly discuss the origin of the giant MR in kagome metal.
For $B_x<1$T, where $\w_H\tau<1$, the MR is proportional to $B_x^2$;
$\rho_{zz}(B_x)=\rho_{zz}(0)+[-\Delta\s_{zz}^\a/\s_{zz}-(\s_{zy}^{{\rm norm},\a})^2/\s_{zz}\s_{yy}]B_x^2$.
The magnetoconductivity $\Delta\s_{zz}^\a$ is given in Eq. (\ref{eqn:Dsigma}).
Here, $\Delta\s_{zz}^\a \propto \sum_{a,\k} f_a' ( \{ d^{x,z}\}^2)_{a,\k}\gamma^{-3}$ and $d^{x,z}=v^y v^{zz}-v^z v^{yz}$.
The $((v^{zz})^2)^+$-term in the integrand gives rise to the giant enhancement of the $\rho_{zz}$ in the $2\times2$ LC state, 
due to the QM term $(g^{zz,+}_a)^2\sim G_a^{zz,+}(v_{ab}^{z,+}v_{ba}^{z,+})$.
Extended Data Fig. \ref{fig:figS5} shows the obtained $\Delta \s_{zz}^x$ of order $O(B^2)$ as function of $\eta$ ($\gamma=1$).
$\Delta \s_{zz}^x$ is enlarged for $\eta\lesssim0.01$ by the $\eta$-even QM.
However, $\Delta \s_{zz}^x$ is already large without the QM,
so the QM-induced enhancement is only quantitative.
In CsV$_3$Sb$_5$, $\rho_{zz}(B_x)$ does not saturate in the quantum limit $\w_H\tau\ll1$ due to the open-orbit Fermi surface along the $k_z$ axis 
\cite{eMChA}.
Nearly $B_x$-linear MR for $B_x\gg1$ would originate from Fermi surfaces with sharp corners in the $2\times2$ LC phase
\cite{AbrikosovS}.

\section{Numerical results of $v^{zz,\pm}_\k$, ${\tilde v}^{zz,\pm}_\k$, $g^{zz,\pm}_\k$, $((g^{zz})^2)^{\pm}_\k$ and $G^{zz,\pm}_\k$}

In this section, we perform the numerical study for $X^{\pm}_\k\equiv \sum_b X^{\pm}_{b,\k}(-f'(\e_{b,\k}^{0}))$, where $X=v^{zz}$, ${\tilde v}^{zz}$, $g^{zz}$, $(g^{zz})^2$, and $G^{zz}$.
Here, $f(\e)$ is Fermi distribution function. 
Note that $f'(0)=-1/4T$.
Below, we set $\eta=0.024$ and $T=0.01$.

\begin{figure}[htb]
\includegraphics[width=.99\linewidth]{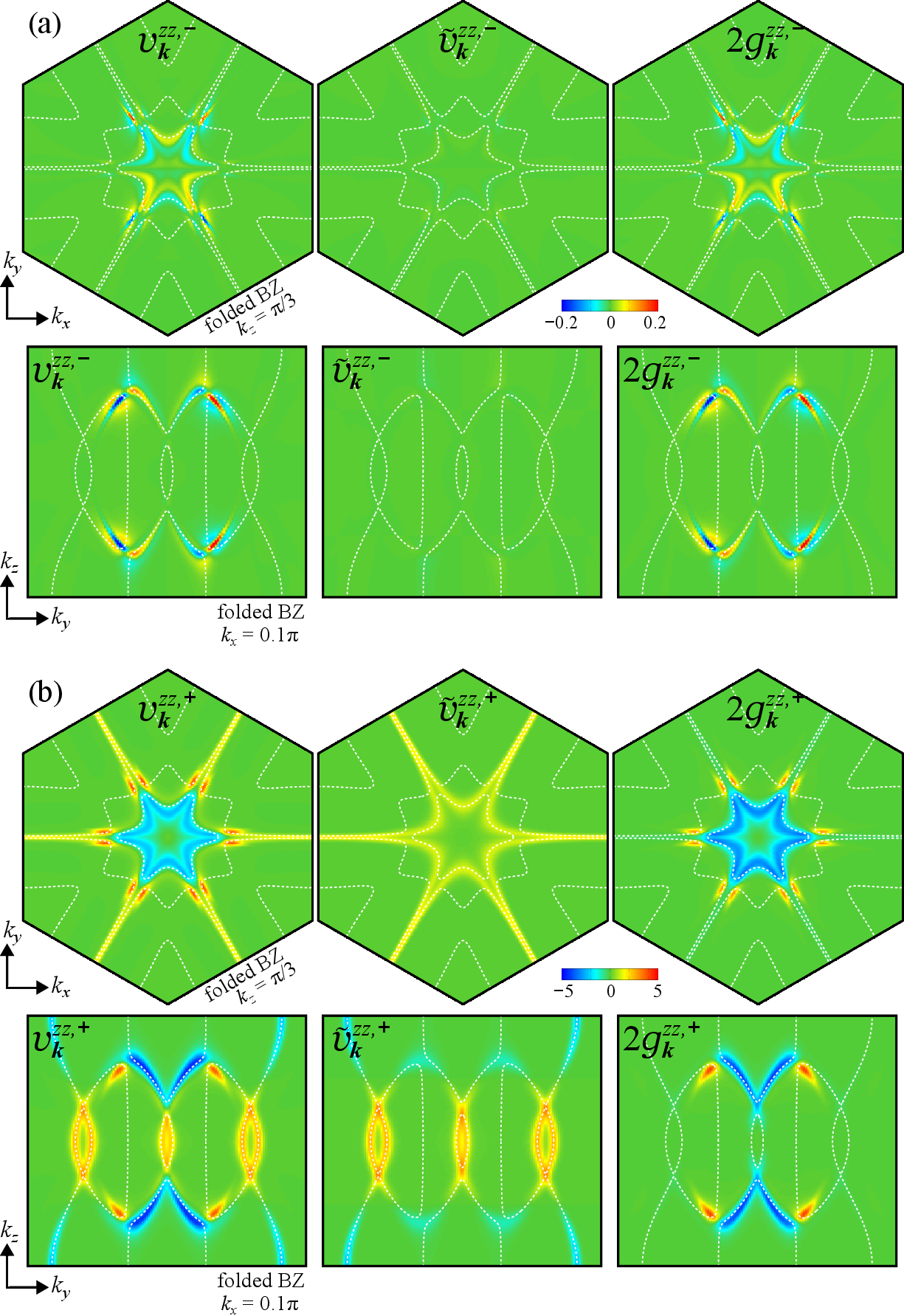}
\caption{
{Results of $v^{zz,\pm}_\k$, ${\tilde v}^{zz,\pm}_\k$, and $g^{zz,\pm}_\k$}: \
(a) Results of $v^{zz,-}_\k$, ${\tilde v}^{zz,-}_\k$, and $g^{zz,-}_\k$ 
around the Fermi level, in the $k_z=\pi/3$ plane and $k_x=0.1\pi$ plane.
The broken lines represent the Fermi surfaces.  
The relation $v^{zz,-}_\k\approx g^{zz,-}_\k$ near the resonant positions is verified. 
(b) Results of $v^{zz,+}_\k$, ${\tilde v}^{zz,+}_\k$, and $g^{zz,+}_\k$.
}
\label{fig:figS6}
\end{figure}

In Extended Data Fig. \ref{fig:figS6} (a), we show the $\eta$-odd functions
$v^{zz,-}_\k$, ${\tilde v}^{zz,-}_\k$ and $g^{zz,\pm}_\k$.
We find that both $v^{zz,-}_\k$ and $g^{zz,\pm}_\k$ exhibit large values with sudden sign changes around the resonant position.
Here, the relation $v^{zz,-}\approx g^{zz,-}$ is obtained
because ${\tilde v}^{zz,-}$ is a small and smooth function of $\k$.
The qQM term $g^{zz,-}$ takes large values near the resonant positions, where the pure-mix band reconstruction occurs, due to the mixture of the LC order ($\eta$) in the pure-band and the large $k_z$-direction dispersion in the mix-band.
As we have explained in the main text, the NLH conductivity is $\s_{yzz}\propto \sum_\k g^{zz,-}_\k v_\k^{y,+} f'$.
Therefore, the NLH is strongly enlarged by the qQM, as we have explained in the main text.


We also show the $\eta$-even functions $v^{zz,+}_\k$, ${\tilde v}^{zz,+}_\k$ and $g^{zz,+}_\k$ in Extended Data Fig. \ref{fig:figS6} (b).
Here, $v^{zz,+}$ deviates from $g^{zz,+}$ for wide area on the Fermi surfaces because ${\tilde v}^{zz,+}$ and $g^{zz,+}$ are comparable in magnitude.
Both $v^{zz,+}$ and $g^{zz,+}$ take large values mainly on the mix-band due to the large $k_z$-direction dispersion of the mix-band.
(The pure-mix hybridization is not necessary for the enhancement of $v^{zz,+}$ and $g^{zz,+}$ because they are of order $O(\eta^0)$.)

\begin{figure}[htb]
\includegraphics[width=.9\linewidth]{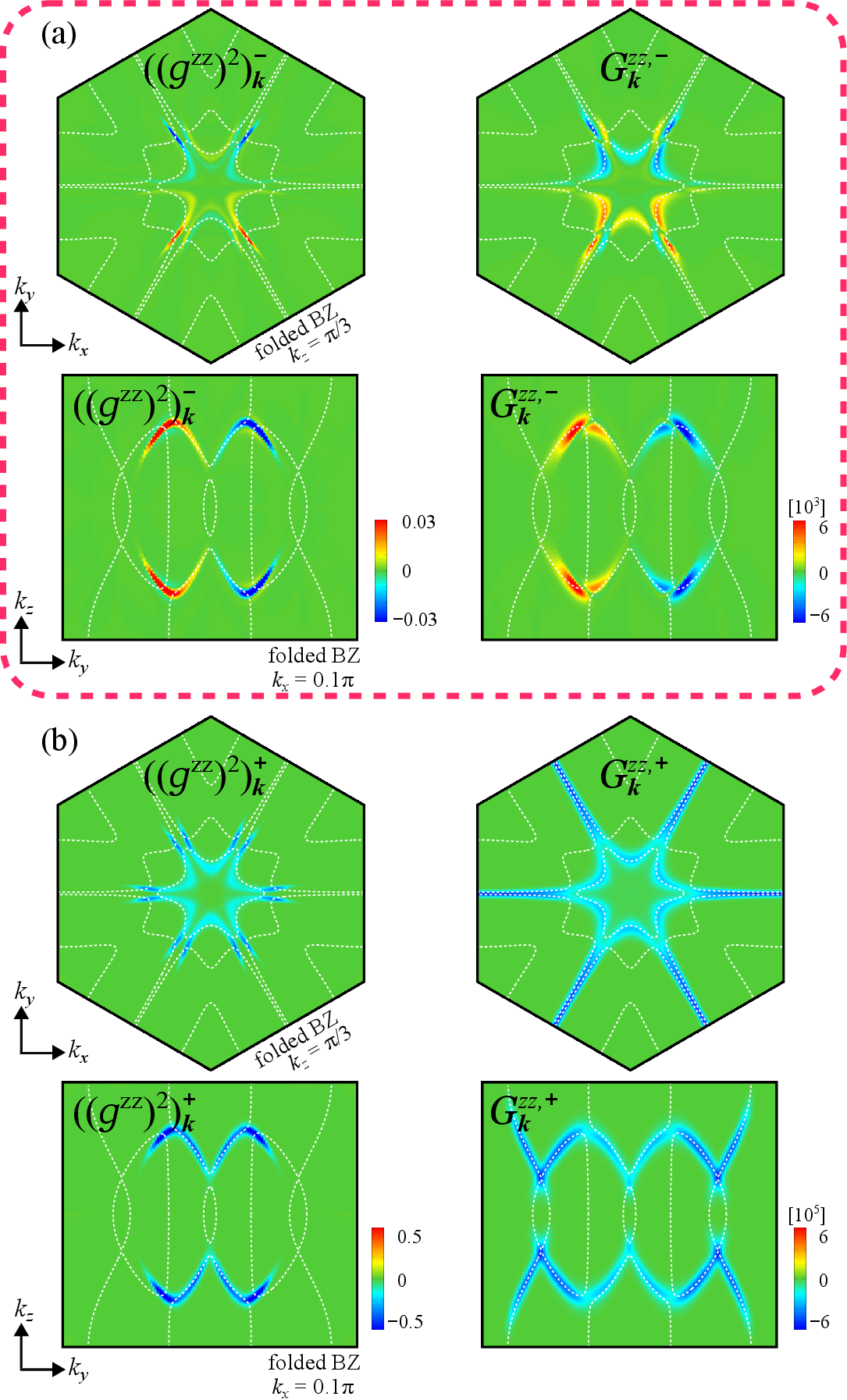}
\caption{
{Results of $((g^{zz})^2)^{\pm}_\k$ and $G^{zz,\pm}_\k$}: \
(a) Results of $((g^{zz})^2)^{-}_\k$ and $G^{zz,-}_\k$
around the Fermi level, in the $k_z=\pi/3$ plane and $k_x=0.1\pi$ plane.
The broken lines represent the Fermi surfaces. 
Here, the relation $((g^{zz})^2)^{-}_\k\propto G^{zz,-}_\k$ near the resonant positions is verified numerically.
This relation is significant for the QM-induced giant eMChA. 
(b) Results of $((g^{zz})^2)^{+}_\k$ and $G^{zz,+}_\k$.
}
\label{fig:figS7}
\end{figure}

In Extended Data Fig. \ref{fig:figS7} (a), we show $\eta$-odd functions $((g^{zz})^2)^{-}_\k$ and $G^{zz,-}_\k$ at $\eta=0.024$ and $T=0.01$.
These quantities exhibit drastic enhancement around the resonant positions. 
Both $((g^{zz})^2)^{-}_\k$ and $G^{zz,-}_\k$ are essentially positive for $k_y<0$ and negative for $k_y>0$.
(In contrast, $g^{zz,-}$ exhibits the sudden sigh change within the area $k_y<0$ (or $k_y>0$); see in Extended Data Fig. \ref{fig:figS6} (a).)
Thus, the relation $((g^{zz})^2)^{-}_\k\propto G^{zz,-}_\k\cdot (t_{{yz}\perp})^2$ has been confirmed numerically.
This relation is easily obtained in the two-band approximation.
The QM term $G^{zz,-}_\k$ take large values near the resonant positions, where the pure-mix band reconstruction occurs, due to the coexistence between the LC order ($\eta$) in the pure-band and the large $k_z$-direction dispersion in the mix-band.
As we have explained in the main text, the eMChA conductivity is $\s_{zzz,x}\propto \sum_\k ((g^{zz})^2)^{-}_\k v_\k^{y,+} f' \propto \sum_\k G^{zz,-}_\k v_\k^{y,+} f'$, where $v_\k^{y,+}\propto k_y$.
Therefore, it is concluded that the giant eMChA originates from the odd-$\eta$ part of the QM.

We also show the $\eta$-even functions $((g^{zz})^2)^{+}_\k$ and $G^{zz,+}_\k$ in Extended Data Fig. \ref{fig:figS7} (b).
Here, $((g^{zz})^2)^{+}$ deviates from $G^{zz,+}$ for wide area on the Fermi surfaces.
Both $((g^{zz})^2)^{+}$ and $G^{zz,+}$ take large values mainly on the mix-band due to the large $k_z$-direction dispersion of the mix-band.
(Note that the pure-band with two-dimensional character is unimportant for $((g^{zz})^2)^{+}$ and $G^{zz,+}$ because they are of order $O(\eta^0)$.)




\end{document}